\documentclass[useAMS,usenatbib]{mn2e}
\usepackage{amsfonts,amssymb}
\usepackage{graphicx}
\usepackage{ulem}
\begin{document}

\title[]{Population Synthesis of Millisecond X-ray Pulsars}
\author[ Zhu et al. ]{Chunhua Zhu$^{1}$, Guoliang L\"{u}$^{1}$\thanks{E-mail:
guolianglv@xao.ac.cn (LGL)}, Zhaojun Wang$^{1}$\\
$^1$School of Physics, Xinjiang University, Urumqi, 830046,
China\\}

\date{\today}




\pagerange{\pageref{firstpage}--\pageref{lastpage}} \pubyear{}

\maketitle

\label{firstpage}

\begin{abstract}
As the evolutionary link between the radio millisecond pulsars (MSPs) and the low mass X-ray binaries or
intermediate mass X-ray binaries, the millisecond X-ray pulsars (MSXPs) are important objects
in testing theories of pulsar formation and evolution. In general, neutron stars in MSXPs can form via core collapse supernova (CC channel) of massive
stars or accretion induced collapse (AIC channel) of an accreting ONeMg WD whose mass reaches the Chandrasekhar limit.
Here, in addition to CC and AIC channels we also consider another channel,
i.e., evolution induced collapse (EIC  channel) of a helium star with mass between $1.4$ and $2.5 M_{\odot}$.
Using a population synthesis code, we have studied MSXPs arising from three different evolutionary channels.
We find that the Galactic birthrates of transient MSXPs
and persistent MSXPs are about 0.7---$1.4\times 10^{-4}$ yr$^{-1}$.
Our population synthesis calculations have shown that about 50\%---90\% of the MSXPs have undergone CC channel,
about 10\%---40\% of them have undergone EIC  channel,
and the MSXPs via AIC channel are the least.

\end{abstract}

\section{Introduction}
Millisecond pulsars (MSPs) are a population of old neutron stars (NSs) which have been
detected in the radio, X-ray and gamma ray wavelengths of the electromagnetic spectrum.
Most of them are in binary systems. Depending on the nature of the companion star MSPs
are believed to form from either low mass X-ray binaries (LMXBs) or intermediate mass
X-ray binaries (IMXBs) \citep[ i.e. recycling scenario,][]{Bhattacharya1991, Tauris2011}.
In this standard scenario, the NS accrete the matter from the companion star,
and its spin frequency becomes very high on a much longer timescale because of angular momentum conservation.
During this process we can detect the system as an X-ray source and towards the end of this process as a millisecond X-ray pulsar (MSXP).
MSXPs
are divided into two subclasses: accretion powered MSXPs and nuclear powered MSXPs.
The former is an X-ray pulsar powered
by accreted material releasing gravitational energy, and the latter is
one powered by accreted material burning in an unstable thermonuclear flash.
On observations, there are type I bursts in nuclear powered MSXPs. However, these two
classes overlap, burst oscillations were observed also from eight accretion powered MSXPs\citep{Papitto2014}.
Not only accretion powered MSXPs but also nuclear powered MSXPs must accrete matter from their companions.
In our work, considering that both of them include accreting NSs, we do not distinguish them,
and call accretion and nuclear powered MSXPs as MSXPs.
Once the companion star can not fill its Roche lobe and the mass transfer terminates, the NS
is observed as a recycled radio MSP. This evolutionary scenario has been supported by the very
recent discovery of "a transition between a rotation powered and an accretion powered state in a
binary MSP" \citep{Papitto2013}.

\begin{figure}
\includegraphics[totalheight=3.0in,width=2.5in,angle=-90]{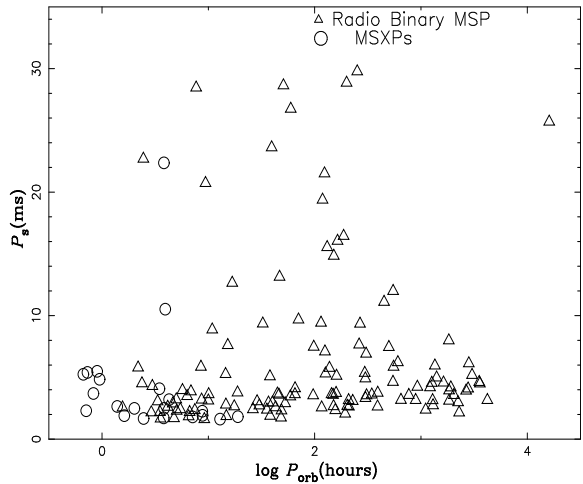}
\caption{The distributions of spin periods ($P_{\rm s}$) and orbital periods ($P_{\rm orb}$) of
MSXPs (circles) and radio binary MSPs (triples). The observational data of the former come from \citet{Patruno2012}
and \citet{Watts2008}, and the data of the latter come from ATNF pulsar catalogue.}
\label{fig:obse}
\end{figure}

As for 5 Aug 2014, about 200 MSPs in the field  and 120 MSPs in globular cluster were known\footnote{https://apatruno.wordpress.com/about/millisecond-pulsar-catalogue/}.
There are 14 accretion powered MSXPs in \cite{Patruno2012}, 16 nuclear powered MSXPs in \cite{Watts2008}, respectively.
\cite{Papitto2013} found a MSP (IGR J18245-2452) which was transiting from accretion to rotation powered emission.
This system also includes an accreting MSP. In our work, MSXPs are the binary systems including accreting MSPs.
Therefore, only 31 MSXPs are observed.
Figure \ref{fig:obse} shows the distribution
of spin periods ($P_{\rm s}$) and orbital periods ($P_{\rm orb}$) of MSXPs
and radio binary MSPs. If MSXPs are the progenitors of radio binary MSPs, we have to explain the followings:
(i)Where do the radio binary MSPs with orbital periods longer than 100 hours come from? (ii) Where do
the MSXPs with orbital periods shorter than 1 hour evolve to?

 Usually, the NS is formed via core collapse (CC) supernova (SN) of a massive star,
 and it receives additional high kick velocity ($\approx 100$ km ${\rm s}^{-1}$).
 Many literatures found that the evolutionary scenario from LMXBs or IMXBs to
 MSPs via CC channel cannot explain the origin of binary MSPs whose orbital periods
 are longer than 100 days \cite[e. g.,][]{Pfahl2003,Liu2011}.
 \cite{Michel1987} considered that an ONeMg white dwarf (WD) accretes sufficient material
 from its companion, and can form a rapidly spinning NS via accretion induced collapse (AIC)
 when its mass reaches the Chandrasekhar limit.
 Considering angular
 momentum conservation, it is possible that accreting ONeMg WD directly collapses into
 MSP via the AIC scenario \citep{Bhattacharya1991}. Assumed that the AIC scenario
 results in a low kick velocity ($\approx 10$ km ${\rm s}^{-1}$), \cite{Hurley2010}
 demonstrated that the AIC channel cannot be ignored in radio binary MSPs formations
 and that some radio binary MSPs in wide orbits (up to about 1000 days) are best
 explained by AIC scenario \citep[Also see][]{Smedley2014}. \cite{Chen2011}
 investigated whether eccentric binary MSPs can form via AIC channel.
 However, both of them did not consider MSXPs.

In the above literatures, authors did not consider that NS can be formed via the third scenario.
\cite{Miyaji1980} considered that ONe core of single star can collapse to a NS by electron capture on $^{20}$Ne and $^{24}$Mg.
This process is called as ECSN.
Because ECSN is largely very dependent on stellar composition, mass-loss rate or internal mixing, \cite{Eldridge2004}
pointed out that the mass range of stars undergoing it is largely uncertain.
For solar metallicity, \cite{Poelarends2007} suggested that the initial-mass range of
stars between $\sim$ 9.0 and $9.2 M_\odot$ can undergo ECSNe.
It is between $\sim$ 9.8 and 9.9 $M_odot$ in \cite{Doherty2015}.
For a single star, the ECSN channel is limited to a very narrow initial mass range.
Its contribution to the total gravitational collapse SN is negligible.

\cite{Podsiadlowski2004} discussed the binary evolution examples. The star in a binary system may experience a process of mass transfer, and it losses whole convective envelope prior to central helium ignition and becomes a naked helium star. Due to the lack of hydrogen-rich convective envelope, the star does not undergo the second dredge-up, which results in a more massive helium core.  Based on \cite{Nomoto1984} and \cite{Nomoto1987}, \cite{Podsiadlowski2004} considered that the naked helium stars with masses between 1.4 and 2.5 $M_\odot$ undergo ECSN.  \cite{Podsiadlowski2004} pointed out that these naked helium stars correspond to initial masses of single stars about 8 --- 11 $M_\odot$. Therefore,
comparing to a single star, the mass range of a star in a binary system for ECSNe occurring is large.
Following \cite{Ivanova2008}, this channel is called as the EIC  scenario. In their work,
NSs in binary systems can be formed via three
scenarios: CC, AIC and EIC. Considering dynamical events (collisions, mergers and so on),
\cite{Ivanova2008} investigated the formation and evolution of LMXBs and MSPs in globular clusters,
and found that AIC and EIC  are important formation channels for NSs in globular clusters. Including single and
binary systems, \cite{Kiel2008} simulated the evolution of the Galactic radio pulsars. Compared with
the observed diagram of spin period vs. spin period derivative for MSPs, they found that low kick velocity
for AIC and EIC  can form a greater number of MSPs. Both of them focused on radio MSPs population in
binary systems. Up to now, there is no detailed theoretical work to
simulate how MSXPs form via the above three scenarios.

Recently, MSXPs have attracted the attention of astronomers. Many X-ray and gamma ray space missions
like RXTE, XMM-Newton, INTEGRAL, Chandra, Swift and HETE
have discovered many observational properties of these objects, such as the orbital periods, spin
period, X-ray luminosities\citep{Watts2008, Altamirano2010, Patruno2012, Papitto2013}.
To our knowledge, there is no investigation of population synthesis for MSXPs to date.
As the evolutionary link between radio MSPs and NSs in LMXBs/IMXBs, MSXPs have played
an important role in testing theories of pulsar formation and evolution. In addition,
the studies of MSXPs will deepen our knowledge about the accretion physics and the properties of NSs.
In this sense, it is necessary to perform a detailed population synthesis of MSXPs.

In this work, we concentrate on MSXPs, and attempt to investigate
the properties of MSXPs formed via the three evolutionary channels.
Section 2 presents our assumptions and describes some details of the modelling algorithm,
and Section 3 gives the main results and discussed in detail. Our main conclusions are summarized in Section 4.

\section{Models}
\cite{Hurley2002} gave the rapid binary star evolution (BSE) code which is often used in many
population synthesis codes. \cite{Kiel2006} updated BSE code. In our work,
we use BSE code in which there are many input parameters. If we do not specially
mention any input parameter, its default value is taken from \cite{Kiel2006}.

The common envelope (CE) evolution and the NS formation and evolution are very important factors
in our simulation. They have great effects on the population of MSXPs. The following three subsections give descriptions.

\subsection{CE Evolution}
In binary systems, CE evolution is very important. Unfortunately, the understanding about it is still very poor. Usually, people considered that dynamically unstable mass transfer results in CE evolution.  In BSE code, there are two input parameters to model CE evolution. The first parameter is the criterion for dynamically unstable mass transfer, $q_{\rm c}$. If the mass ration of the components ($q=M_{\rm donor}/M_{\rm gainer}$) is larger than $q_{\rm c}$, the binary system experiences CE evolution. However, the value about $q_{\rm c}$ is still uncertain \citep{Han2001,Han2002}. Considering non-conservative Roche lobe overflow, \cite{Han2001} found that $q_{\rm c}$ may be up to about 1.3 \citep[also see][]{Chen2008}. Very recently, \cite{Pavlovskii2015} showed that $q_{\rm c}$ can varies from 1.5 to 2.2 if the recombination energy in the superadiabatic layer of giant donors was considered.
In addition, \cite{King1999} and \cite{Tauris2000} suggested that the CE evolution is probably avoided in any binary in which the convective envelope of the donor star is not too deep if accretion rate is super-Eddington.
In order to discuss the effects of $q_{\rm c}$ on MSXPs¡¯ population, we simulate three cases as the follows:
(i)Following \cite{Hurley2002}, we take $q_{\rm c}$=3 (0.695) if the donor is a main sequence (MS) star with mass larger (lower) than 0.7 $M_\odot$, respectively. If the donor is a giant we take $q_{\rm c}$ as \citep{Webbink1988}
\begin{equation}
q_{\rm c}=0.362+\frac{1}{3\times(1-M_{\rm c}/M_{\rm donor})},
\label{eq:qcrit}
\end{equation}
where $M_{\rm c}$ is the core mass of the donor;\\
(ii)For giant donor, $q_{\rm c}=1.5$(\cite{Han2001, Chen2008, Pavlovskii2014}), while other $q_{\rm c}$s are the same with (i);\\
(iii)CE evolution only occurs when the donor in a binary is a giant \citep{King1999,Tauris2000}. Therefore, for giant donor, $q_{\rm c}$=Eq(\ref{eq:qcrit}), while in order to avoid CE evolution other $q_{\rm c}$s are equals to 100.

The second parameter is the efficiency of the binary orbital energy used to expel the envelope of donor, $\alpha_{\rm ce}$.
 However, based on a balance equation for the orbital angular momentum of binary system, \cite{Nelemans2000} introduced another parameter $\gamma$ to describe CE evolution. The former is called as $\alpha$-algorithm, and the later is called as $\gamma$-algorithm. Similarly, the value of $\gamma$ is also very uncertain. In order to discuss the effect of CE evolution on MSXPs¡¯ population, we carry out different simulations in this work. In $\alpha$-algorithm, $\alpha_{\rm ce}\lambda_{\rm ce}=1.0$, and $\gamma=1.5$ in $\gamma$-algorithm. Here $\lambda_{\rm ce}$ is a stellar structure parameter,  and it depends on the evolutionary stage of the donor. As \cite{Nelemans2000} suggested, the $\gamma$-algorithm can be applied only in the first CE evolution for binary systems in which two low-mass stars have similar masses. Considering that CE evolution is crucial in close binaries, we use $\gamma$-algorithm to simulate the first CE evolutions for all binary systems. When binary systems undergo the second CE evolutions, we use $\alpha$-algorithm.

\subsection{Formation Channels of NSs and Kick Velocity}
\label{sec:channel}
As described in the introduction, there are three channels for the formation of NSs in binary systems \citep{Ivanova2008, Kiel2008}:  (i) CC channel for a star with MS mass $ M/M_\odot \geq 11$; (ii) AIC channel for an accreting ONeMg WD when its mass reaches the Chandrasekhar limit; (iii) EIC  channel for naked  helium stars with  masses between $1.4$ and $2.5 M_{\odot}$. In BSE code, these naked helium stars originate from initially more massive stars in binary systems with masses in the range of 8 --- 11 $M_\odot$ which is consistent with the range of stellar masses quoted by \cite{Podsiadlowski2004}.
 It is questionable if an MSP would form directly from any kind of channel \citep{Tauris2013}. In our simulations all NSs formed via three channels need an accretion phase and are subsequently spun up to become MSXPs. The reasons are discussed in section 3.3.2

The evolution of accreting ONeMg WD in AIC channel is poorly known. For simplicity, we use the same way of calculating the
evolution of a CO WD by accretion of hydrogen-rich material
from its companion. The amount of matter retained by accreting WD is critical for success of AIC channel.
For the retention efficiencies for hydrogen accumulation ($\eta_{\rm H}$), we follow \cite{Hachisu1999} (also see \cite{Han2004} and \cite{Lu2009}).
Based on the model of optically thick wind, \cite{Kato2004} calculated the retention efficiencies for helium accumulation ($\eta_{\rm He}$).
Considering theoretical and observational evidence against existence of optically thick winds from accreting WD \citep{Lepo2013},
we used $\eta_{\rm He}$ calculated by \cite{Piersanti2014} who took into account existence of Roche lobes in binaries
instead of assuming optically thick winds. However, in order to discuss the effects of the retention efficiencies on MSXPs¡¯ population, we also used Kato-Hachisu retention efficiencies in a case. In addition, from the work of \cite{Jose1998}, it is known that ONe WDs have about 2 times higher retention efficiencies. Therefore, we carried out two simulations in which the retention efficiencies of accreting ONe WDs were 2 or 1/2 times of those calculated by \cite{Piersanti2014}.

The limits of stable accumulating matter also determine the mass growth of accreting WDs.
The limits for accreting CO WDs depend on the WD¡¯s masses \citep[e. g.,][]{Hachisu1999,Kato2004, Piersanti2014}, and the larger is CO WD¡¯s mass, the higher is the limit. Usually, compared to CO WDs, ONe WDs are massive. Therefore, in our work, we extrapolated the limits for CO WDs to ONe WDs although they may be difference.

Not only observations but also theoretical simulations, it is very difficult to give a certain value for kick velocity received by nascent NS. \cite{Pfahl2002} suggested that the nascent NSs via CC channel have high kick velocity ($\sim 100\,\rm km\ s^{-1}$), while the NSs born in EIC  and AIC have low kick velocity($\sim 10\,\rm km\ s^{-1}$). Like in our study about the donors of NSs in persistent LMXBs \citep{Zhu2012}, the distribution of kick velocity ($v_{\rm k}$) is given by
\begin{equation}
P(v_{\rm k})=\sqrt{\frac{2}{\pi}}\frac{v^2_{\rm k}}{\sigma^3_{\rm
k}}e^{-v^2_{\rm k}/2\sigma^2_{\rm k}}.
\end{equation}
\cite{Hobbs2005} investigated the proper motion of 233 pulsars, and found that a Maxwellian distribution with $\sigma_{\rm k}= 265$ km s$^{-1}$  can describe well the velocity distribution of these pulsars. However, in order to check the effect of $v_{\rm k}$ on MSXPs¡¯ population, we take $\sigma_{\rm k}= 190$ and 400 km s$^{-1}$ for CC channel \citep{Hansen1997}, while $\sigma_{\rm k}^{*}= 20$ and 50 km s$^{-1}$ for EIC  and AIC channels in different simulations, respectively.

\subsection{Evolutions of NSs¡¯ Spin and Magnetic Field}
The evolutions of NS¡¯s spin and magnetic field are mainly determined by the interaction of the NS with surrounding materials. This interaction has become the most important but still is an open problem. For the evolution of NS¡¯s spin, we use a convenient way elaborated by \cite{Lipunov1992}.
Spin evolution (spin-up or spin-down) of a NS in a binary system can be conveniently described by an angular momentum conservation equation
\begin{equation}
\frac{{\rm d}I\omega}{{\rm d}t}=K_{\rm su}-K_{\rm sd}\,,
\label{eq:torq}
\end{equation}
where $I$ is  NS momentum of inertia, $K_{\rm su}$ and $K_{\rm sd}$ are spin-up and spin-down torques, respectively. We treat the NS's spin
evolutions in the same way
as in our previous study for symbiotic X-ray binaries\citep{Lu2012}. The values of spin-up and
spin-down torques exerted on a rotating magnetized NS are summarized in Table 2 of \cite{Lu2012}. In this work, the value of the index n,
which reflects different rotational distributions of
matter inside the shell, is taken 2 \citep{Shakura2012}(for further details see \citet{Lu2012}).

There is no an exact mechanism to describe the decay of NS¡¯s magnetic field. Following \cite{Lu2012}, we assume, for an accreting NS, the decay of magnetic field depends exponentially on the amount of accreted matter. We use the formulae suggested by \cite{Oslowski2011}. For a non-accreting NS, we assume that the decay of magnetic field results from Ohmic decay\citep{Goldreich1992}, and the decay formulae is given by Eq. (2) in \cite{Kiel2008}.
Following \cite{Faucher2006} and \cite{Popov2010},  a log-normal distribution of the initial magnetic fields of nascent NSs is assumed.
The median value of the distribution is 12 and its dispersion equals 1.

In LMXBs/IMXBs, it is possible that not all matter from their companions via Roche lobe flows or stellar winds is accreted by NSs. We introduce a parameter $\beta$ which is the fraction of transferred matter accreted by the NS, and the rest of the transferred matter is lost from binary system. The lost matter takes away the specific angular momentum of the gainer. The value of $\beta$ has been usually set to 0.5 \citep{Podsiadlowski1992,Podsiadlowski2002,Nelson2003}. In our work, in order to compare the influence of $\beta$, we set $\beta=0.25$ and 1.0 in different simulations. In many systems, the mass transfer rates may exceed the Eddington limit. However, not all of materials are accreted. In this work, we assume that a star can accrete material up to the Eddington limit, and any excess material is lost from the binary system.

\subsection{Persistent and Transient X-ray Sources}
On observation, the MSXPs can be classified into persistent
and transient X-ray sources. In all 31 known MSXPs,
there are 24 transient sources and 7 persistent sources\citep{Patruno2012,Watts2008,Papitto2013}.
In this paper, we consider the thermal disk instability
and make a distinction between persistent and transient sources.
The thermal disk instability depends on the mass-accretion rate of NS. The X-ray binary is a transient source when the mass-accretion rate is lower than a certain critical value, $\dot{M}_{\rm crit}$. That is, if $\dot{M}_{\rm NS}<\dot{M}_{\rm crit}$ in Roche
overflow accretion, the system is a transient source, here $\dot{M}_{\rm NS}$ is the mass-accretion rate of NS.
If $\dot{M}_{\rm NS}>\dot{M}_{\rm crit}$ or wind-fed accretion, the
system is a persistent source.
According to \citet{Belczynski2008}, for the hydrogen-rich and heavier-element disks we use the works of \cite{Dubus1999} and \cite{Menou2002}, respectively:
\begin{equation}
\dot{M}_{\rm crit}=\left\{
\begin{array}{ll}
 1.5\times10^{15}M_{\rm NS}^{-0.4}R_{\rm
 d}^{2.1}C^{-0.5}_{1}\  {\rm g/s}, &{\rm H\ rich}\\
 5.9\times10^{16}M_{\rm NS}^{-0.87}R_{\rm
 d}^{2.62}\alpha^{0.44}_{0.1}\  {\rm g/s}, &{\rm He\ rich}\\
 1.2\times10^{16}M_{\rm NS}^{-0.74}R_{\rm
 d}^{2.21}\alpha^{0.42}_{0.1}\ \ \  {\rm g/s}, &{\rm C\ rich}\\
 5.0\times10^{16}M_{\rm NS}^{-0.68}R_{\rm
 d}^{2.05}\alpha^{0.45}_{0.1}\ \ \  {\rm g/s}, &{\rm O\ rich}\\
\end{array}
\right. \label{eq:nhydr}
\end{equation}
where $M_{\rm NS}$ is the NS mass in solar mass, $R_{\rm d}$ is the maximum disk radius (it is approximately equal to 2/3 of accretor Roche lobe radius) in $10^{10}$ cm, constant $C_{1}=C/(5\times10^{-4})$ in which $C$ is a radiation parameter and its
typical value is $5\times10^{-4}$, $\alpha_{0.1}=\alpha/0.1$ in which
$\alpha=0.1$ is a viscosity parameter.

\section{Population of MSXPs}
Using the method of population synthesis for binary population, several initial input parameters (initial mass function, initial mass ration of binary components and initial binary separation) must be given. For the primaries in binary systems, their masses are given by the initial mass function of \cite{Miller1979}. For initial mass ration of binary components, we adopt a flat distribution \citep{Kraicheva1989, Goldberg1994}. Following \cite{Yungelson1993}, we assume that initial binary separation, $A_0$, is between 10 and $10^6 R_\odot$, and has a flat distribution over $\log\ A_0$.

In this work, we considered eight cases in which different input parameters are listed in Table \ref{tab:case}.
In order to investigate the birthrates of MSXPs,
we assume simply a constant star formation rate over last 13 Gyr and that there is one binary with $M_1 \geq 0.8 M_\odot$\ born per year in the Galaxy\citep{Iben1984,Yungelson1993,Han1995a}.
In every case, we simulate the evolutions of $2\times10^8$ binary systems.
This gives a statistical error between about 1\% and 2\% for the birthrate of the Galactic MSXPs via different channels.

\subsection{Birthrates and Average Lifetime}
The birthrates and the average lifetime of the transient MSXPs and the persistent MSXPs with
different kinds of donors in the Galaxy are summarized in Tables \ref{tab:result1} and \ref{tab:result2}, respectively.

\begin{table}
\centering
  \caption{Parameters of the models for MSXPs' populations. The first column gives model number.
           Column 2 shows the algorithm of CE evolution. Column 3 gives the dispersion of
           kick-velocity distribution for CC channel, and column 4 for AIC and EIC channels.
           The critical mass ratio for dynamically unstable mass transfer, $q_{\rm c}$ is given in
           column 5, in which (i), (ii) and (iii) are described in subsection 2.1.
           Column 6 shows the retention efficiencies for helium accumulation ($\eta_{\rm He}$),
           in which P and K-$\eta_{\rm He}$ are calculated by \citet{Piersanti2014} and \citet{Kato2004},
           respectively.}
  \tabcolsep1.0mm
 \begin{tabular*}{85mm}{cccccc}
\hline \hline
\multicolumn{1}{c}{Cases}&\multicolumn{1}{c}{CE}&\multicolumn{1}{c}{$\sigma_{\rm k}$(km
s$^{-1}$)}&\multicolumn{1}{c}{$\sigma_{\rm k}^{*}$(km
s$^{-1}$)}&\multicolumn{1}{c}{$q_{\rm c}$}&\multicolumn{1}{c}{$\eta_{\rm He}$}\\
 \cline{1-6}
Case 1&$\alpha_{\rm ce}\lambda_{\rm
ce}=1.0$&190&20&(i)&P-$\eta_{\rm He}$\\
Case 2&$\alpha_{\rm ce}\lambda_{\rm
ce}=1.0$&400&50&(i)&P-$\eta_{\rm He}$\\
Case 3&$\gamma=1.5$&190&20&(i)&P-$\eta_{\rm He}$\\
Case 4&$\alpha_{\rm ce}\lambda_{\rm
ce}=1.0$&190&20&(ii)&P-$\eta_{\rm He}$\\
Case 5&$\alpha_{\rm ce}\lambda_{\rm
ce}=1.0$&190&20&(iii)&P-$\eta_{\rm He}$\\
Case 6&$\alpha_{\rm ce}\lambda_{\rm
ce}=1.0$&190&20&(i)&K-$\eta_{\rm He}$\\
Case 7&$\alpha_{\rm ce}\lambda_{\rm
ce}=1.0$&190&20&(i)&$2\times$P-$\eta_{\rm He}$\\
Case 8&$\alpha_{\rm ce}\lambda_{\rm
ce}=1.0$&190&20&(i)&$\frac{1}{2}\times$P-$\eta_{\rm He}$\\

 \cline{1-6}
 \label{tab:case}
\end{tabular*}
\end{table}

\begin{table*}
\centering
 \begin{minipage}{170mm}
  \caption{Different models of the transient MSXPs' population. The first column gives the model number according to Table \ref{tab:case}.
  Columns 2 to 7 give the birthrates and the average lifetime of MSXPs with different
  kinds of donors, respectively.  Total birthrate and lifetime are
  showed in columns 8 and 9, respectively. NS + MS means that accreting NS has a main sequence donor,  NS + RG means that accreting NS has a red giant donor,
  NS + WD represents that accreting NS has a WD donor, and '---' means that we do not get any sample in $2\times10^8$ binary systems.}
  \tabcolsep1.7mm
  \begin{tabular*}{170mm}{ccccccccc}
\hline\hline
\multicolumn{1}{c}{}&\multicolumn{2}{c}{NS+MS}&
\multicolumn{2}{c}{NS+RG}&
\multicolumn{2}{c}{NS+WD}&
\multicolumn{2}{c}{Total}\\
\multicolumn{1}{c}{Cases}&\multicolumn{1}{c}{Birthrate}
&\multicolumn{1}{c}{Lifetime}&\multicolumn{1}{c}{Birthrate}
&\multicolumn{1}{c}{Lifetime}&\multicolumn{1}{c}{Birthrate}
&\multicolumn{1}{c}{Lifetime}&\multicolumn{1}{c}{Birthrate}
&\multicolumn{1}{c}{Lifetime}\\
\multicolumn{1}{c}{}&\multicolumn{1}{c}{($\rm yr^{-1}$)}
&\multicolumn{1}{c}{($\rm yr$)}&\multicolumn{1}{c}{($\rm yr^{-1}$)}
&\multicolumn{1}{c}{($\rm yr$)}&\multicolumn{1}{c}{($\rm yr^{-1}$)}
&\multicolumn{1}{c}{($\rm yr$)}&\multicolumn{1}{c}{($\rm yr^{-1}$)}
&\multicolumn{1}{c}{($\rm yr$)}\\
(1)&(2)&(3)&(4)&(5)&(6)&(7)&(8)&(9)\\
\hline
\multicolumn{1}{c}{}&\multicolumn{8}{c}{CC}\\
Cases 1, 6---8\ \ \ \ &3.5$\times 10^{-5}$ & 2.5$\times 10^{9}$&1.6$\times 10^{-5}$ & 7.0$\times 10^{8}$&1.3$\times 10^{-5}$ & 1.2$\times 10^{8}$
&6.4$\times 10^{-5}$ & 2.3$\times 10^{9}$\\
Case 2\ \ \ \ &1.7$\times 10^{-5}$ & 2.7$\times 10^{9}$ &2.5$\times 10^{-6}$ & 7.3$\times 10^{8}$&1.6$\times 10^{-6}$ & 1.9$\times 10^{8}$&2.1$\times 10^{-5}$ & 2.6$\times 10^{9}$\\
Case 3\ \ \ \ &5.9$\times 10^{-5}$ & 3.3$\times 10^{9}$ &9.4$\times 10^{-6}$ & 3.1$\times 10^{8}$&5.2$\times 10^{-6}$ & 2.1$\times 10^{8}$&7.4$\times 10^{-5}$ & 3.2$\times 10^{9}$\\
Case 4\ \ \ \ &2.0$\times 10^{-5}$ & 2.9$\times 10^{9}$ &1.7$\times 10^{-5}$ & 1.0$\times 10^{8}$&9.0$\times 10^{-6}$ & 1.0$\times 10^{8}$&4.6$\times 10^{-5}$ & 2.8$\times 10^{9}$\\
Case 5\ \ \ \ &4.4$\times 10^{-5}$ & 2.5$\times 10^{9}$ &2.0$\times 10^{-5}$ & 5.3$\times 10^{8}$&2.1$\times 10^{-5}$ & 9.0$\times 10^{7}$&8.5$\times 10^{-5}$ & 2.3$\times 10^{9}$\\




\hline
\multicolumn{1}{c}{}&\multicolumn{8}{c}{AIC}\\

Case 1\ \ \ \ & --- \ \ \ & --- \ \ \ & 1.6$\times
10^{-7}$ \ \ \ & 8.1$\times 10^{6}$ \ \ \ & 7.2$\times 10^{-6}$ \ \ \ & 8.1$\times
10^{8}$ \ \ \ & 7.4$\times 10^{-6}$ \ \ \ & 8.1$\times
10^{8}$ \\
Case 2\ \ \ \ & --- \ \ \ & --- \ \ \ & 3.3$\times
10^{-7}$ \ \ \ & 4.5$\times 10^{6}$ \ \ \ & 1.4$\times 10^{-5}$ \ \ \ & 3.8$\times
10^{8}$ \ \ \ &1.4$\times 10^{-5}$ \ \ \ & 3.8$\times
10^{8}$ \\
Case 3\ \ \ \ &--- \ \ \ & --- \ \ \ & 1.3$\times
10^{-6}$ \ \ \ & 2.0$\times 10^{7}$ \ \ \ & --- \ \ \ & --- \ \ \ &1.3$\times 10^{-6}$ \ \ \ & 2.0$\times
10^{7}$  \\
Case 4\ \ \ \ & --- \ \ \ & --- \ \ \ & 7.6$\times
10^{-6}$ \ \ \ & 1.6$\times 10^{6}$ \ \ \ & 4.5$\times 10^{-6}$ \ \ \ & 1.4$\times
10^{8}$ \ \ \ &1.2$\times 10^{-5}$ \ \ \ & 1.4$\times
10^{8}$ \\
Case 5\ \ \ \ &--- \ \ \ & --- \ \ \ & 7.3$\times
10^{-7}$ \ \ \ & 8.5$\times 10^{6}$ \ \ \ & 1.3$\times 10^{-5}$ \ \ \ & 8.0$\times
10^{7}$ \ \ \ &1.4$\times 10^{-5}$ \ \ \ & 8.0$\times
10^{7}$  \\
Case 6\ \ \ \ & 4.8$\times 10^{-7}$ \ \ \ & 4.3$\times 10^{9}$ \ \ \ & 5.4$\times
10^{-7}$ \ \ \ & 9.0$\times 10^{7}$ \ \ \ & 1.4$\times 10^{-5}$ \ \ \ & 5.8$\times
10^{8}$ \ \ \ &1.5$\times 10^{-5}$ \ \ \ & 1.3$\times
10^{9}$  \\
Case 7\ \ \ \ & 1.6$\times 10^{-7}$ \ \ \ & 3.6$\times 10^{9}$ \ \ \ & 4.4$\times
10^{-7}$ \ \ \ & 1.7$\times 10^{7}$ \ \ \ & 1.4$\times 10^{-5}$ \ \ \ & 3.8$\times
10^{8}$ \ \ \ &1.5$\times 10^{-5}$ \ \ \ & 6.9$\times
10^{8}$\\
Case 8\ \ \ \ & --- \ \ \ & ---\ \ \ & 1.0$\times
10^{-7}$ \ \ \ & 3.2$\times 10^{6}$ \ \ \ & 1.4$\times 10^{-6}$ \ \ \ & 3.8$\times
10^{8}$ \ \ \ &1.5$\times 10^{-6}$ \ \ \ & 3.8$\times
10^{8}$  \\
\hline
\multicolumn{1}{c}{}&\multicolumn{8}{c}{EIC}\\
Case 1, 6---8\ \ \ \ & 1.8$\times 10^{-5}$ \ \ \ & 1.3$\times 10^{9}$ \ \ \ & 9.0$\times
10^{-6}$ \ \ \ & 6.2$\times 10^{8}$ \ \ \ & 1.2$\times 10^{-5}$ \ \ \ & 4.4$\times
10^{8}$ \ \ \ & 3.9$\times 10^{-5}$ \ \ \ & 1.1$\times
10^{9}$ \\
Case 2\ \ \ \ & 2.5$\times 10^{-5}$ \ \ \ & 1.3$\times 10^{9}$ \ \ \ & 3.7$\times
10^{-6}$ \ \ \ & 1.4$\times 10^{9}$ \ \ \ & 4.9$\times 10^{-6}$ \ \ \ & 4.5$\times
10^{8}$ \ \ \ &  3.4$\times 10^{-5}$ \ \ \ & 1.3$\times
10^{9}$\\
Case 3\ \ \ \ & 2.8$\times 10^{-6}$ \ \ \ & 1.0$\times 10^{9}$ \ \ \ & 1.6$\times
10^{-5}$ \ \ \ & 2.9$\times 10^{8}$ \ \ \ & 4.7$\times 10^{-5}$ \ \ \ & 6.1$\times
10^{8}$ \ \ \ &  6.0$\times 10^{-5}$ \ \ \ & 6.0$\times
10^{8}$\\
Case 4\ \ \ \ & 1.8$\times 10^{-5}$ \ \ \ & 1.0$\times 10^{9}$ \ \ \ & 8.7$\times
10^{-6}$ \ \ \ & 4.4$\times 10^{8}$ \ \ \ & 1.2$\times 10^{-5}$ \ \ \ & 3.6$\times
10^{8}$ \ \ \ & 3.9$\times 10^{-5}$ \ \ \ & 8.1$\times
10^{8}$ \\
Case 5\ \ \ \ & 1.9$\times 10^{-5}$ \ \ \ & 1.0$\times 10^{9}$ \ \ \ & 1.2$\times
10^{-5}$ \ \ \ & 4.6$\times 10^{8}$ \ \ \ & 1.6$\times 10^{-5}$ \ \ \ & 3.9$\times
10^{8}$ \ \ \ & 4.7$\times 10^{-5}$ \ \ \ & 7.8$\times
10^{8}$\\


\hline \label{tab:result1}
\end{tabular*}
\end{minipage}
\end{table*}

\begin{table*}
\centering
 \begin{minipage}{170mm}
  \caption{Similar to Table \ref{tab:result1}, but for the persistent MSXPs' population. }
  \tabcolsep1.7mm
  \begin{tabular*}{170mm}{ccccccccc}
\hline\hline
\multicolumn{1}{c}{}&\multicolumn{2}{c}{NS+MS}&
\multicolumn{2}{c}{NS+RG}&
\multicolumn{2}{c}{NS+WD}&
\multicolumn{2}{c}{Total}\\
\multicolumn{1}{c}{Cases}&\multicolumn{1}{c}{Birthrate}
&\multicolumn{1}{c}{Lifetime}&\multicolumn{1}{c}{Birthrate}
&\multicolumn{1}{c}{Lifetime}&\multicolumn{1}{c}{Birthrate}
&\multicolumn{1}{c}{Lifetime}&\multicolumn{1}{c}{Birthrate}
&\multicolumn{1}{c}{Lifetime}\\
\multicolumn{1}{c}{}&\multicolumn{1}{c}{($\rm yr^{-1}$)}
&\multicolumn{1}{c}{($\rm yr$)}&\multicolumn{1}{c}{($\rm yr^{-1}$)}
&\multicolumn{1}{c}{($\rm yr$)}&\multicolumn{1}{c}{($\rm yr^{-1}$)}
&\multicolumn{1}{c}{($\rm yr$)}&\multicolumn{1}{c}{($\rm yr^{-1}$)}
&\multicolumn{1}{c}{($\rm yr$)}\\
(1)&(2)&(3)&(4)&(5)&(6)&(7)&(8)&(9)\\
\hline
\multicolumn{1}{c}{}&\multicolumn{8}{c}{CC}\\
Case 1, 6---8\ \ \ \ & 2.8$\times 10^{-5}$ \ \ \ & 4.5$\times 10^{6}$ \ \ \ & 4.0$\times
10^{-6}$ \ \ \ & 1.1$\times 10^{6}$ \ \ \ & 1.6$\times 10^{-5}$ \ \ \ & 1.2$\times
10^{5}$ \ \ \ &4.8$\times 10^{-5}$ \ \ \ & 4.3$\times
10^{6}$ \\

Case 2\ \ \ \ & 1.6$\times 10^{-5}$ \ \ \ & 5.1$\times 10^{6}$ \ \ \ & 1.0$\times
10^{-6}$ \ \ \ & 1.1$\times 10^{6}$ \ \ \ & 2.7$\times 10^{-6}$ \ \ \ & 1.3$\times
10^{5}$ \ \ \ &2.0$\times 10^{-5}$ \ \ \ & 5.0$\times
10^{6}$  \\

Case 3\ \ \ \ & 4.5$\times 10^{-5}$ \ \ \ & 4.8$\times 10^{6}$ \ \ \ & 5.5$\times
10^{-6}$ \ \ \ & 5.3$\times 10^{5}$ \ \ \ & 8.7$\times 10^{-6}$ \ \ \ & 1.3$\times
10^{5}$ \ \ \ &5.9$\times 10^{-5}$ \ \ \ & 4.7$\times
10^{6}$\\

Case 4\ \ \ \ & 1.5$\times 10^{-5}$ \ \ \ & 5.7$\times 10^{6}$ \ \ \ & 8.2$\times
10^{-6}$ \ \ \ & 1.5$\times 10^{6}$ \ \ \ & 2.8$\times 10^{-5}$ \ \ \ & 1.3$\times
10^{5}$ \ \ \ &5.1$\times 10^{-5}$ \ \ \ & 5.0$\times
10^{6}$\\

Case 5\ \ \ \ & 3.4$\times 10^{-5}$ \ \ \ & 4.7$\times 10^{6}$ \ \ \ & 7.1$\times
10^{-6}$ \ \ \ & 9.7$\times 10^{5}$ \ \ \ & 1.6$\times 10^{-5}$ \ \ \ & 1.2$\times
10^{5}$ \ \ \ & 5.7$\times 10^{-5}$ \ \ \ & 4.5$\times
10^{6}$ \\

\hline
\multicolumn{1}{c}{}&\multicolumn{8}{c}{AIC}\\
Case 1\ \ \ \ & --- \ \ \ & --- \ \ \ & 1.5$\times
10^{-7}$ \ \ \ & 4.8$\times 10^{5}$ \ \ \ & 4.5$\times 10^{-6}$ \ \ \ & 1.2$\times
10^{5}$ \ \ \ & 4.6$\times 10^{-6}$ \ \ \ & 1.6$\times
10^{5}$\\
Case 2\ \ \ \ & --- \ \ \ & --- \ \ \ & 3.1$\times
10^{-7}$ \ \ \ & 2.7$\times 10^{5}$ \ \ \ & 8.5$\times 10^{-6}$ \ \ \ & 1.2$\times
10^{5}$ \ \ \ &  8.8$\times 10^{-6}$ \ \ \ & 1.3$\times
10^{5}$\\
Case 3\ \ \ \ &--- \ \ \ & --- \ \ \ & 1.3$\times
10^{-6}$ \ \ \ & 3.1$\times 10^{5}$ \ \ \ & --- \ \ \ & --- \ \ \ & 1.3$\times 10^{-6}$ \ \ \ & 3.1$\times
10^{5}$\\
Case 4\ \ \ \ & --- \ \ \ & --- \ \ \ & 1.7$\times
10^{-6}$ \ \ \ & 1.4$\times 10^{5}$ \ \ \ & 2.3$\times 10^{-6}$ \ \ \ & 1.2$\times
10^{5}$ \ \ \ &  4.6$\times 10^{-6}$ \ \ \ & 1.3$\times
10^{5}$\\
Case 5\ \ \ \ &--- \ \ \ & --- \ \ \ & 7.0$\times
10^{-7}$ \ \ \ & 4.7$\times 10^{5}$ \ \ \ & 6.7$\times 10^{-6}$ \ \ \ & 1.2$\times
10^{5}$ \ \ \ &  7.4$\times 10^{-6}$ \ \ \ & 2.2$\times
10^{5}$\\
Case 6\ \ \ \ & 4.8$\times 10^{-7}$ \ \ \ & 3.7$\times 10^{6}$ \ \ \ & 4.9$\times
10^{-7}$ \ \ \ & 6.6$\times 10^{5}$ \ \ \ & 8.6$\times 10^{-6}$ \ \ \ & 1.2$\times
10^{5}$ \ \ \ &  9.6$\times 10^{-6}$ \ \ \ & 2.2$\times
10^{6}$\\
Case 7\ \ \ \ & 1.6$\times 10^{-7}$ \ \ \ & 3.7$\times 10^{6}$ \ \ \ & 4.2$\times
10^{-7}$ \ \ \ & 5.9$\times 10^{5}$ \ \ \ & 8.5$\times 10^{-6}$ \ \ \ & 1.2$\times
10^{5}$ \ \ \ &9.1$\times 10^{-6}$ \ \ \ & 1.3$\times
10^{6}$ \\
Case 8\ \ \ \ & --- \ \ \ & ---\ \ \ & 1.0$\times
10^{-7}$ \ \ \ & 3.9$\times 10^{5}$ \ \ \ & 8.5$\times 10^{-7}$ \ \ \ & 1.2$\times
10^{5}$ \ \ \ &  9.5$\times 10^{-7}$ \ \ \ & 1.6$\times
10^{5}$\\

\hline
\multicolumn{1}{c}{}&\multicolumn{8}{c}{EIC}\\
Case 1, 6---8\ \ \ \ & 5.6$\times 10^{-5}$ \ \ \ & 2.3$\times 10^{6}$ \ \ \ & 1.2$\times
10^{-5}$ \ \ \ & 2.8$\times 10^{5}$ \ \ \ & 1.1$\times 10^{-5}$ \ \ \ & 1.9$\times
10^{5}$ \ \ \ &7.9$\times 10^{-5}$ \ \ \ & 2.2$\times
10^{6}$ \\

Case 2\ \ \ \ & 8.9$\times 10^{-5}$ \ \ \ & 1.6$\times 10^{6}$ \ \ \ & 3.6$\times
10^{-6}$ \ \ \ & 2.2$\times 10^{5}$ \ \ \ & 4.1$\times 10^{-9}$ \ \ \ & 2.1$\times
10^{5}$ \ \ \ &9.8$\times 10^{-5}$ \ \ \ & 1.6$\times
10^{6}$\\

Case 3\ \ \ \ & 5.6$\times 10^{-6}$ \ \ \ & 4.1$\times 10^{6}$ \ \ \ & 1.8$\times
10^{-5}$ \ \ \ & 2.8$\times 10^{5}$ \ \ \ & 3.8$\times 10^{-5}$ \ \ \ & 1.5$\times
10^{5}$ \ \ \ &  6.2$\times 10^{-5}$ \ \ \ & 2.9$\times
10^{6}$\\

Case 4\ \ \ \ & 5.6$\times 10^{-5}$ \ \ \ & 2.3$\times 10^{6}$ \ \ \ & 1.2$\times
10^{-5}$ \ \ \ & 2.8$\times 10^{5}$ \ \ \ & 1.1$\times 10^{-5}$ \ \ \ & 1.9$\times
10^{5}$ \ \ \ &  2.9$\times 10^{-5}$ \ \ \ & 1.7$\times
10^{6}$\\

Case 5\ \ \ \ & 5.8$\times 10^{-5}$ \ \ \ & 2.3$\times 10^{6}$ \ \ \ & 2.0$\times
10^{-5}$ \ \ \ & 2.1$\times 10^{5}$ \ \ \ & 1.4$\times 10^{-5}$ \ \ \ & 1.9$\times
10^{5}$ \ \ \ & 4.0$\times 10^{-5}$ \ \ \ & 1.6$\times
10^{6}$\\


\hline \label{tab:result2}
\end{tabular*}
\end{minipage}
\end{table*}

In our simulations, the total birthrate of the transient MSXPs arising
from the three channels is from 6.9 $\times 10^{-5}$ (Case 2)
to 1.4 $\times 10^{-4} \rm yr^{-1}$ (Case 3). The total birthrate of the persistent MSXPs
is from 8.4 $\times 10^{-5}$ (Case 4) to 1.3 $\times 10^{-4} \rm yr^{-1}$ (Case 1). Considering
CC and AIC channels,
\cite{Hurley2010} studied radio binary MSPs, and showed that their birthrate is between
5.5 $\times 10^{-4} \rm yr^{-1}$ and  2.2 $\times 10^{-5} \rm yr^{-1}$ in the simulations with different
input parameters. According to the theories of radio MSPs formation and evolution, the radio emission is activated when the companion of NS can not fill up its Roche lobe and the mass transfer terminates. That is, the MSXPs will eventually
evolve to the radio MSPs, which means that the birthrate of the radio MSPs is about $10^{-4}$ ---$10^{-5}$ yr$^{-1}$ in the Galaxy.
 Our results consist with those in \cite{Hurley2010}.
However, the birthrate of the radio MSPs established from their radio properties,
is about 3---5$\times10^{-6}$ yr$^{-1}$ \citep{Lorimer2005,Ferrario2007,Story2007},
 a factor 10---100 times lower than our calculations. This means that we encounter a problem of ¡®overproduction¡¯ of MSXPs.

The  ¡®overproduction¡¯ problem may results from the followings:\\ (i)Selection effects. The sample of observed MSXPs is heavily biased towards
the brighter objects that are the easiest to detect. The observed MSXPs are only the tip of the iceberg of a much larger underlying population.\\
(ii)The calculation of mass-transfer rates. The lifetimes of the MSXPs depend on mass-transfer rates. In our work, we use BSE code to calculate the
mass-transfer rates. Compared to full evolutionary computations, BSE code may give inaccurate results. Using BSE code and
MESA code, \cite{Chen2014} calculated the mass-transfer rates of some binary systems, respectively. They found that BSE code choices lower
mass-transfer rates than that in MESA code when the donors are red giants. Our work may overestimate the lifetimes
of MSXPs.\\
(iii)The X-ray active lifetimes of the transient MSXPs. In general, the transient MSXPs are relatively faint X-ray sources. Their typical quiescent
luminosities are about $10^{32}$--- $10^{33} \ \rm erg \ s^{-1}$ and the peak X-ray luminosities
during outbursts are about $10^{36} \ \rm erg \ s^{-1}$ \citep{Papitto2011, Paizis2012}. During the quiescent phase,
it is very difficult to observe the transient MSXPs as X-ray sources. The duty cycles (the fraction
of time during the outburst) are very uncertain. Empirically the duty cycles is not exceeding $1\%$ \citep{Belczynski2008}.
\cite{Haaften2012} used theoretical approach to calculate duty cycles which depend on the mass,
radius and mass-accretion rate of the gainer and binary separation. They found that the duty cycles
in some ultra-compact X-ray binaries decrease below 0.1\%.
However, \cite{Yan2014} performed a statistical study of
the outburst properties of 110 bright X-ray outbursts in 36 LMXB transients.
They found that the duty cycles of these transients with NSs are about 1\%---10\%, and the average duty cycles are 3.5\%.

A similar problem appears in the study of LMXBs: about 1000 ---10000 strong X-ray systems are predicted by population synthesis codes\citep{Pfahl2003},
while less than 200 LMXBs are observed \citep{Liu2007}.
Therefore, in order to estimate their number in the Galaxy,
we assume that the observed probability of a transient MSXPs approximately equal 3.5\%\citep{Yan2014}.
Then, we estimate that there are about 3700 (case 2) --- 9800 (case 3) observable transient MSXPs in the Galaxy.
According to our calculations, about 53\% (Case 2)---86\% (Case 3)
of the transient MSXPs arise from CC channel, and about 14\% (Case 3)---42\% (Case 2) of them arise from EIC  channel, while the transient MSXPs arising
from AIC channel are less than 3\% and can be negligible because of too short lifetime.

Similarly, based on Table \ref{tab:result2}, there are about 340 (case 2) --- 500 (case 3) persistent MSXPs in the Galaxy. About 30\% (Case 2)---70\% (Case 4)
of the persistent MSXPs arise from CC channel, about 10\% (Case 4)---40\% (Case 2) of them arise from EIC  channel
and about 10\% (Case 3)---30\% (Case 2) of them arise
from AIC channel. In our work, there should be 4000---10000 MSXPs in the Galaxy. However, up to now, there
are only 31 known MSXPs in the Galaxy. Like previous studies on LMXBs, we also overestimate the number of MSXPs.

In our work, an important result must be noted.
Although it is unclear whether all MSXPs become MSPs\citep{Tauris2012},
we still assume that all MSXPs should eventually evolve to radio MSPs.
Taking  Case 1 as an example, we find that the birthrate of the radio MSPs formed via CC channel
is $6.4\times 10^{-5}\ \rm yr^{-1}$ which is consistent with \cite{Hurley2010}. The birthrates of the
radio MSPs via AIC and EIC  channels are $7.4\times 10^{-6}$ and $3.9\times 10^{-5}\ \rm yr^{-1}$, respectively.
Usually, all nascent radio MSPs have low magnetic fields. Their lifetimes as radio sources approximately equal \citep{Ferrario2007}.
This means that about 58\% of radio MSPs result from CC channel, 7\% of radio MSPs result from
AIC channel, and 35\% of radio MSPs result from EIC  channel. Therefore the EIC  channel is one
of the most important channels in the formation of the radio MSPs, and cannot be
ignored in studies of the MSPs.

\subsection{Effects of Parameters}
In our work, different input parameters have effects on the populations of MSXPs.
Compared with Case 1, the parameters $\sigma_k$ and  $\sigma_k^*$ are increased from 190 to 400 $\rm km\ s^{-1}$
and from 20 to 50 $\rm km\ s^{-1}$  in Case 2, respectively.
The larger $\sigma_k$ and  $\sigma_k^*$ are, the more difficultly a binary survives after SN. Therefore, the
birthrate of MSXPs in Case 2 decreases.

In general,  $\alpha_{\rm ce}\lambda_{\rm ce}=1.0$ in $\alpha$-algorithm means that a binary orbital period after CE evolution should
shorten to 1\% ---10\% of the orbital period at beginning of Roche lobe overflow. While, $\gamma=1.5$ in $\gamma$-algorithm means that
a binary orbital period is approximately constant during CE evolution.
In our simulations, different algorithms of CE evolutions have different effects on the MSXPs via different channels.
For CC and EIC  channels, the birthrate of the MSXPs in Case 3 is larger than that of in Case l,
however, for AIC channel, the result is opposite.
The reason is that, for AIC channel, an ONeMg WD should accrete sufficient matter from its companion star
before it collapses to a rapidly spinning NS. The distances of post CE
systems with the $\alpha$-algorithm are narrower than those with the $\gamma$-algorithm, which is favor for ONeMg WDs accreting more matter.
So, for AIC channel, the birthrate of the MSXPs in Case 1 ($\alpha$-algorithm) is larger than that of in Case 3 ($\gamma$-algorithm).

Compared to Case 1, $q_{\rm c}$ is larger for giant donors in Case 4. It means that CE occurs more difficultly in the latter.
 Therefore, there are more NS + RG systems and less NS + WD systems in Case 4. The assumption of $q_{\rm c}$ in Case 5 mainly avoids
 CE evolution when donor is on Hertzsprung Gap, and the material of the donor's envelope transfers to gainer. When the donor evolves into
 giant, its envelope becomes CE. Based CE evolutionary theory, less is the mass of CE, more can binary survive after CE formation.
 In Case 5, there are more MSXPs. According to Cases 1, 4 and 5, $q_{\rm c}$ affects
 the birthrates and lifetimes of MSXPs within a factor of 1.5.

 The MSXPs' birthrate via AIC channel mainly depends on the parameter $\eta_{\rm He}$ (retention efficiency for helium accumulation ) in this work.
 Comparing Cases 1, 6 and 7, we find that the $\eta_{\rm He}$ calculated by \cite{Kato2004} is higher than that by \cite{Piersanti2014},
 and the former is approximately two times as high as the latter. From $\eta_{\rm He}=2\times P$-$\eta_{\rm He}$ in Case 7 to  $\eta_{\rm He}=1/2\times P$-$\eta_{\rm He}$ in Case 8, the birthrate of MSXPs via AIC channel reduces by about 10 times.


\subsection{Properties of MSXPs via the Three Evolutionary
Scenarios}
The present work focuses on the MSXPs populations via CC, AIC and EIC  scenarios. In
this subsection we discuss the properties of MSXPs via the three evolutionary scenarios.
In the following we take Case 1 as an example to discuss some properties (orbital periods,
spin periods, donor's masses, mass-accretion rates) of MSXPs populations.

\subsubsection{Initial Orbital Period and Donor's Mass: $P^{\rm i}_{\rm orb}$ vs. $M^{\rm i}_{2}$}
\begin{figure*}
\centering
\begin{tabular}{c@{\hspace{3pc}}c}
\includegraphics[totalheight=3.0in,width=2.5in,angle=-90]{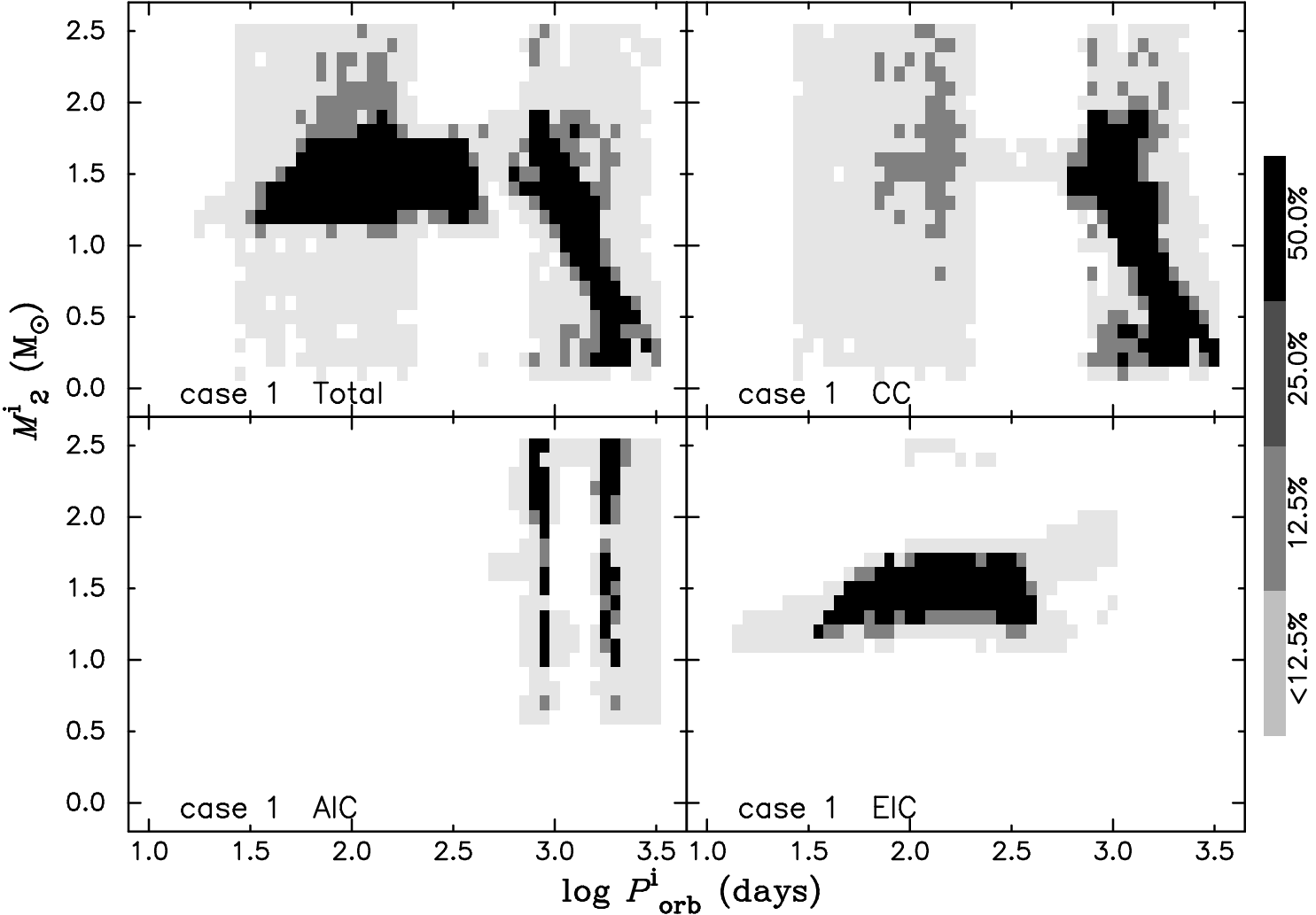}&
\includegraphics[totalheight=3.0in,width=2.5in,angle=-90]{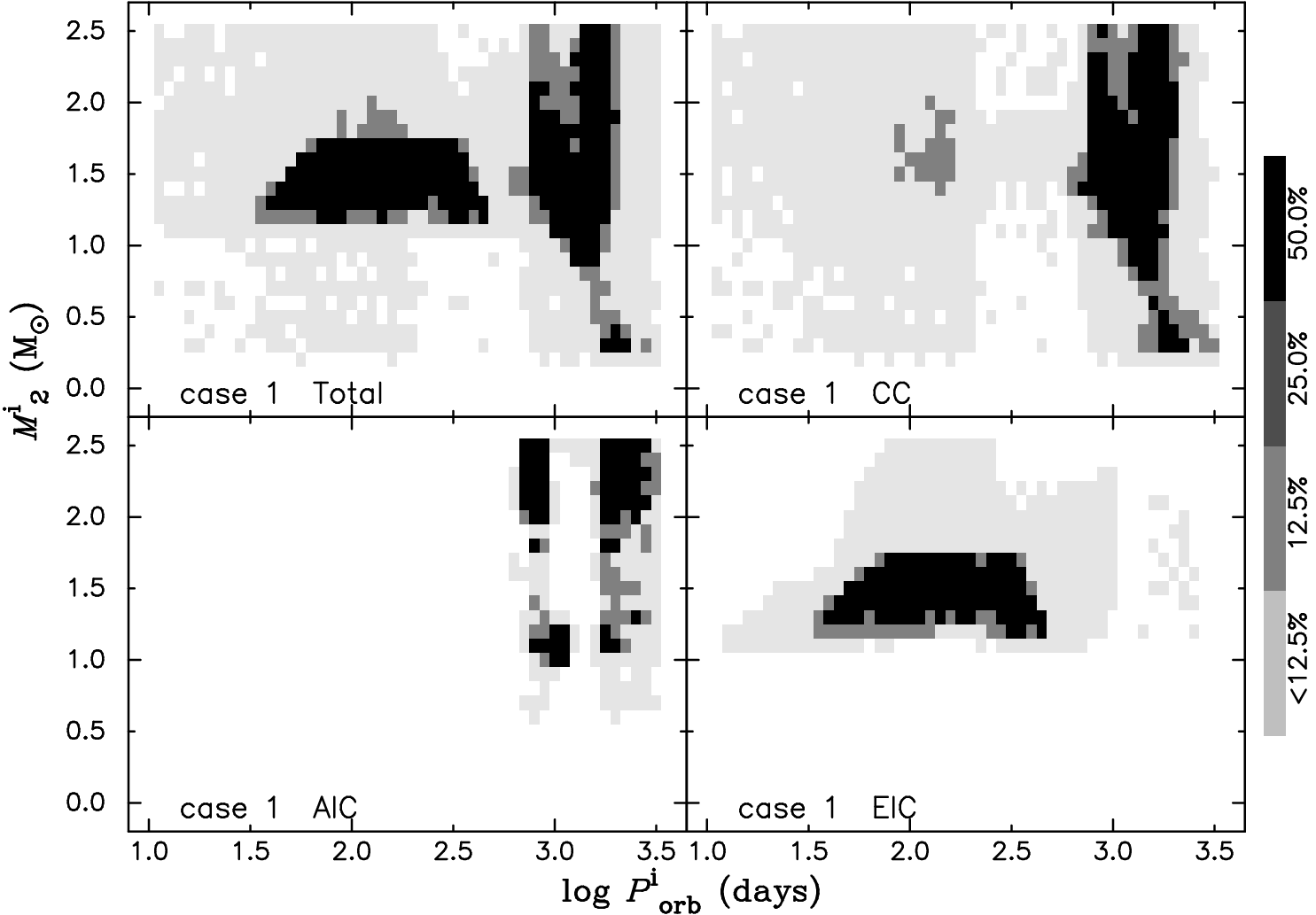}\\
\end{tabular}
\caption{Gray-scale maps of the distributions of the initial orbital periods $P_{\rm orb}$ vs.
the initial masses of accreting NSs' companions in the MSXPs.
The left is for the transient sources, and the right is for the persistent sources.}
\label{fig:im2po}
\end{figure*}

Figure \ref{fig:im2po} shows the distributions of the initial orbital periods $P_{\rm orb}$ vs.
the initial masses of secondaries (They will become the donors of MSXPs) for the progenitors of MSXPs. Compared
with the progenitors of transient sources, the progenitors of persistent sources have larger initial secondaries' masses but
they have similar orbital periods. On average,  among the three evolutionary scenarios,
the progenitors of MSXPs via EIC scenario have the shortest
initial orbital periods because they must undergo CE evolution to form naked helium stars when
the primaries just leave MS stage, while the progenitors of MSXPs via AIC  scenario have the
longest initial orbital periods because the primaries must have enough space to form ONe WDs.

Most of
binary systems in Figure \ref{fig:im2po} undergo violent binary interaction,
and then evolve into MSXPs. Compared with their progenitors, MSXPs have
very shorter orbital periods which are showed in Figure \ref{fig:porb}.
The left graph is for the transient MSXPs, and the right is for the persistent MSXPs.
For the transient sources, there are three different zones for $P_{\rm orb}$s' distributions.
The left zone ($P_{\rm orb}<1$ hour) presents the ultra-compact MSXPs, that is, the donors in these
MSXPs are WDs or naked helium stars. The transient MSXPs with MS donors mainly are in the middle zone ($1<P_{\rm orb}<50$ hours),
and the right zone ($P_{\rm orb}>50$ hour) is mainly composed of the transient MSXPs with red giant donors.
All of three types of donors appear in the transient MSXPs arising three channels.
Obviously, the orbital periods of transient MSXPs
via AIC channel are shortest because in our simulations they undergo twice stable mass transfers: one is that the accreting ONe
WD collapses into NS, another is that the accreting NS is spin-up into MSP. Therefore, in our
work, AIC channel is no contribution to MSXPs with long orbital periods.
\cite{Hurley2010} assumed that
accreting ONe WD can directly collapse into MSP without a spin-up process. In their
work, the contribution of binary MSPs via AIC channel is significant.

\subsubsection{Orbital Period: $P_{\rm orb}$}
\label{sec:pb}
\begin{figure*}
\centering
\begin{tabular}{c@{\hspace{3pc}}c}
\includegraphics[totalheight=3.0in,width=2.5in,angle=-90]{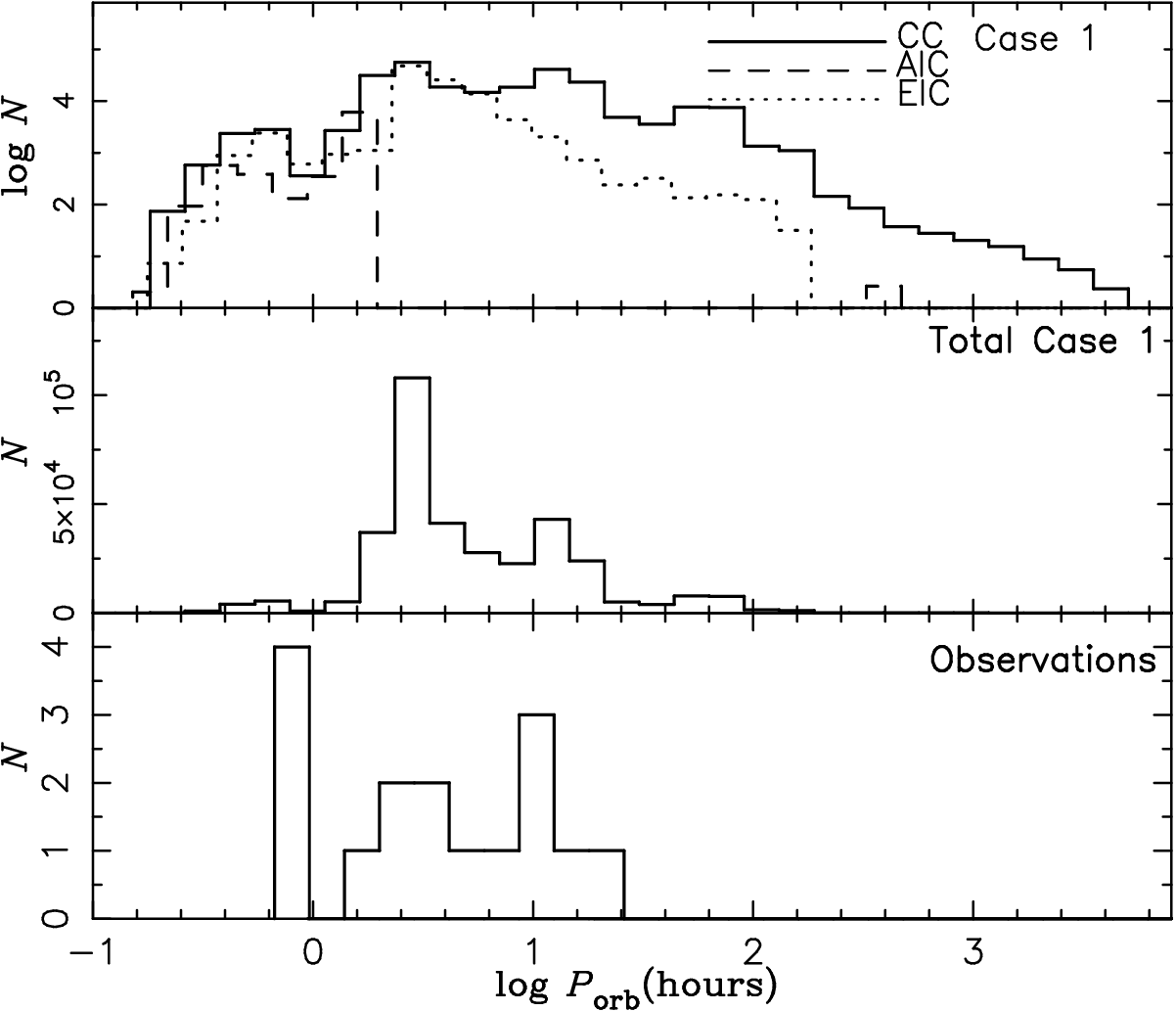}&
\includegraphics[totalheight=3.0in,width=2.5in,angle=-90]{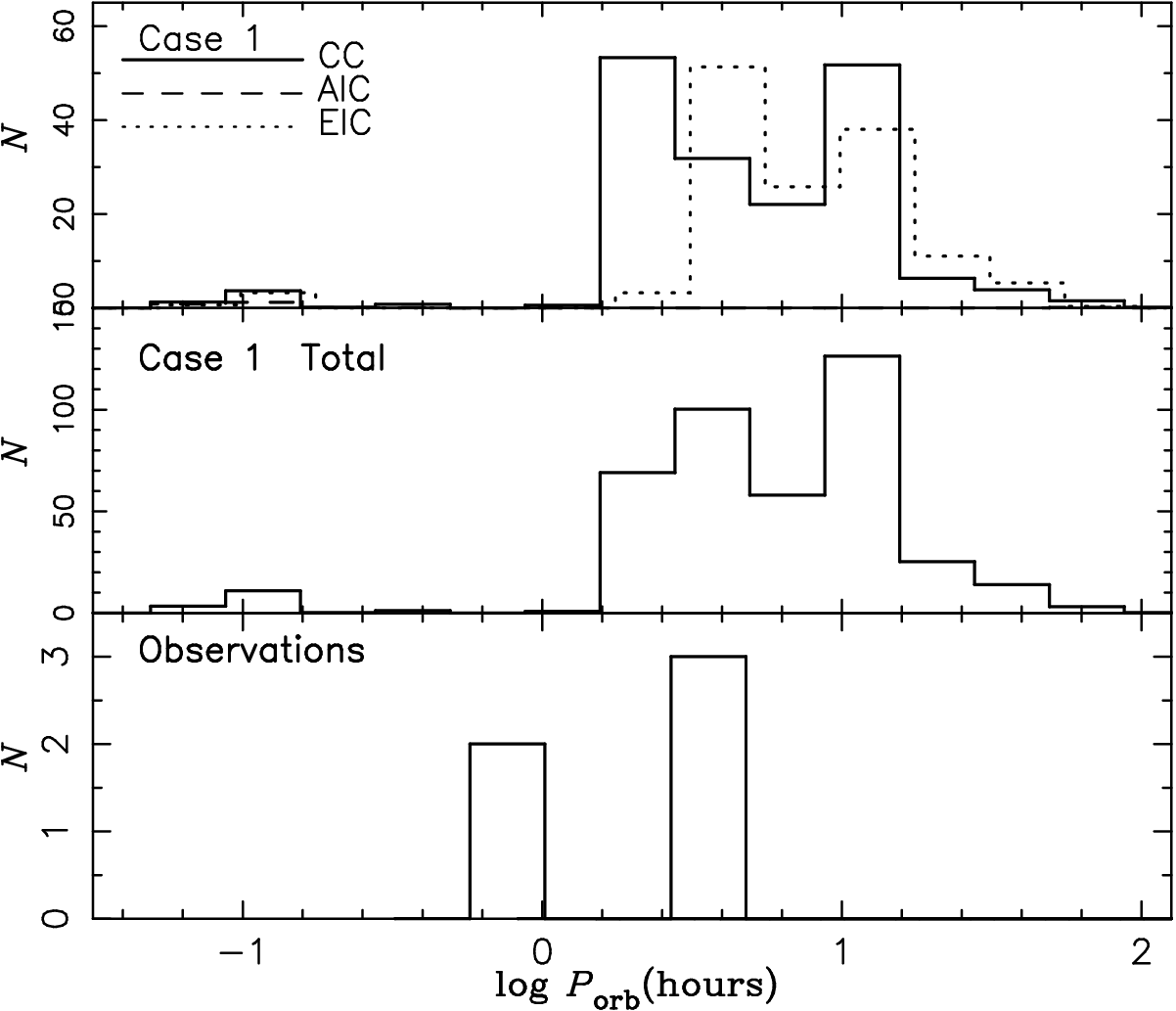}\\
\end{tabular}
\caption{The distributions of the orbital periods of MSXPs. The left is for the transient sources,
and the right is for the persistent sources. The data of observed MSXPs come from \citet{Watts2008, Patruno2012, Papitto2013}.}
\label{fig:porb}
\end{figure*}

Similarly, as the right graph in Figure \ref{fig:porb} and Table \ref{tab:result2} show, the most of
the persistent MSXPs have MS donors, and the ultra-compact persistent MSXPs only is about 2\% of
the whole persistent MSXPs.  Comparing the left graph with the right one in Figure \ref{fig:porb}, the
orbital periods of the transient MSXPs can be wider than those of the persistent MSXPs. The main reason comes from
our definition for the transient and the persistent MSXPs (See Eq. (\ref{eq:nhydr})).

\begin{figure}
\includegraphics[totalheight=3.0in,width=2.5in,angle=-90]{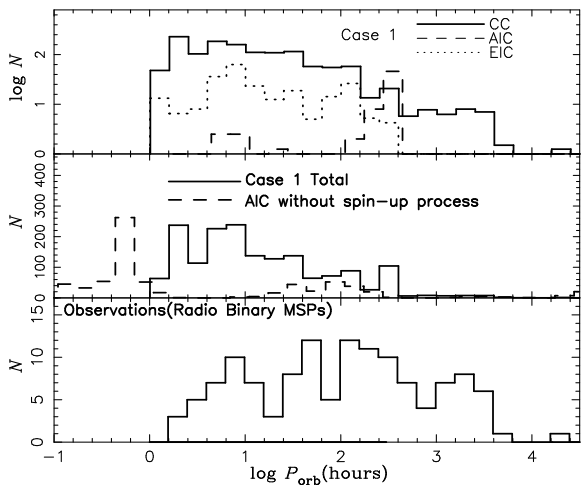}
\caption{The distributions of the orbital periods of MSXPs when they terminate mass transfer
and become radio binary MSPs in Hubble time. The dashed line in the middle panel represents radio binary
MSPs directly formed via AIC channel without spin-up process.
The orbital periods observed in all known radio binary MSPs are plotted in the bottom panel.
The data of observed MSXPs come from ATNF pulsar catalogue.}
\label{fig:onwd}
\end{figure}

In all known MSXPs, the histogram of their orbital periods observed are plotted
in the bottom panels of Figure \ref{fig:porb}.
As the middle and the bottom panels in Figure \ref{fig:porb} show, our results cover all observational samples well.
However, two aspects in our results are worth noting:\\
(i) In all known MSXPs, the orbital period (18.95 hours) of Aql X-1 is the most long. However, according to
Figure \ref{fig:obse}, many radio binary MSPs have orbital periods longer than 100 days. In fact, in our simulation,
there are some MSXPs with orbital periods longer than 100 days. However, the duration of they staying in X-ray stage
is too short. Therefore, they are negligible and hardly are showed in the histogram.
Figure \ref{fig:onwd} shows the distributions of orbital periods of MSXPs when they terminate
mass transfer and become radio binary MSPs.
If all radio binary MSPs originate from MSXPs, it is difficult to use our model to explain the formation of radio binary MSPs with
long orbital period. Similarly done in \cite{Hurley2010}, we also do a
simulation in which we assume that radio binary MSPs can be directly formed via AIC channel without spin-up process. Our
result is plotted by dashed line in the middle panel of Figure \ref{fig:onwd}. Obviously, our results
still do not explain why so many radio binary MSPs have orbital periods longer than 1000 hours. \\
(ii)Based on Tables \ref{tab:result1}, \ref{tab:result2} and Figure \ref{fig:porb},
the ultra-compact transient MSXPs only is about 4\% of the whole transient MSXPs, and
the ultra-compact persistent MSXPs only is about 2\% of the persistent MSXPs.
However, in the known 24 transient MSXPs,
there are 5 ultra-compact transient MSXPs, and in the known 7 persistent MSXPs, there is 1 ultra-compact persistent MSXP.
Obviously, we underestimate
the ratio of the ultra-compact MSXPs to the total MSXPs. The possible causes will be discussed
in the next subsection.

In addition, where do these ultra-compact MSXPs evolve into? \cite{Deloye2003} considered
the effect of ideal gas pressure, degeneracy pressure and Coulomb attraction on the WD radius,
and found that some ultra-compact MSXPs can evolve into radio binary MSPs with orbital periods
70 --- 90 minutes within 5 --- 10 Gyr. Recently, having thought
the propeller effect and disk instability in ultra-compact X-ray binaries, \cite{Haaften2012} considered
that they evolved into low mass ratio binaries with orbital periods 70 --- 80 minutes.
In the ultra-compact MSXPs, we use a zero-temperature WD mass-radius
relation, and  orbital changes mainly originate from gravitational radiation and mass variations. In
our work, with the donor's mass decreases which results in the decrease of mass transfer rate,
the ultra-compact MSXPs evolve into the faint ultra-compact X-ray binaries,
and hardly evolve into radio binary MSPs in Hubble time. Therefore, as Figure \ref{fig:onwd} shows,
there is no radio binary MSP with orbital period shorter than 1 hour in our work. However, if
we assume that MSPs can directly form via AIC channel without spin-up process, there
are some radio binary MSPs with orbital periods shorter than 1 hour (See the dashed line in
the middle panel of Figure \ref{fig:onwd}).  They may evolve into the ultra-compact MSXPs due
to gravitational radiation within 10 Gyr. Therefore,
we should observe some radio binary MSPs with orbital periods shorter than 1 hour. However,
to our knowledge, there is no known radio binary MSP with orbital period shorter than 1 hour.
We considered that there may be two kinds of explanation: \\
(i) If ONe WD can directly collapse into a MSP via AIC channel, the newborn MSP may have
high magnetic field because no accreted material buries magnetic field.   \cite{Paradijs1997} considered that the bursting pulsar GRO J1744-28 originates from a massive ONe WD via AIC \citep[Also see][]{Xu2009}.
Although GRO J1744-28 is not a MSP (its pulsar period is 467 ms),  \cite{Paradijs1997}
assumed that it was a MSP when it was born. According to the observational date of pulse period and spin-up rate, \cite{Finger1996} estimated that the magnetic field of GRO J1744-28 is of order $10^{11}$ G. Therefore,
the newborn MSP via AIC channel may be a magnetic field of $\sim 10^{12}$ G. Treating the pulsar
as a rotating magnetic dipole, the surface magnetic field strength $B=3.2\times10^{19} (P \dot{P})^{1/2}$ G,
and the characteristic age $\tau_{\rm c}=P/(2\dot{P})$. Then, we can estimate that $\tau_{\rm c}\sim 10^6$ yr if the
newborn MSP via AIC channel has a spin period of about $10$ ms. Compared to normal
MSPs with low magnetic field ($B\sim10^{8-9}$ G) and long characteristic age($\tau_{\rm c}\sim10^9$ yr),
the lifetime of the newborn MSP via AIC channel as a radio MSP is very short. Therefore, it is difficult to find them. \\
(ii) The process of WD collapsing into a MSP via AIC channel may be very complex, and the angular
momentum of ONe WD can be lost by some mechanism. For example, the SN via AIC scenario ejects
some matter which can carry out more angular momentum. However, to our knowledge,
there is not any simulation or observation to refer it.\\

Therefore, it is possible that NS via AIC channel can not directly become MSPs, or it is difficult to observed these MSPs even though they
can be directly formed via AIC channel. However,
if the NSs via AIC channel undergo spin-up processes to become MSPs, they should be similar with other MSPs.
Considering the above reasons, we assume that the MSPs via AIC channel must undergo spin-up processes in the present work.

\subsubsection{Orbital Period and Donor's Mass: $P_{\rm orb}$ vs. $M_2$}
\begin{figure*}
\centering
\begin{tabular}{c@{\hspace{3pc}}c}
\includegraphics[totalheight=3.0in,width=2.5in,angle=-90]{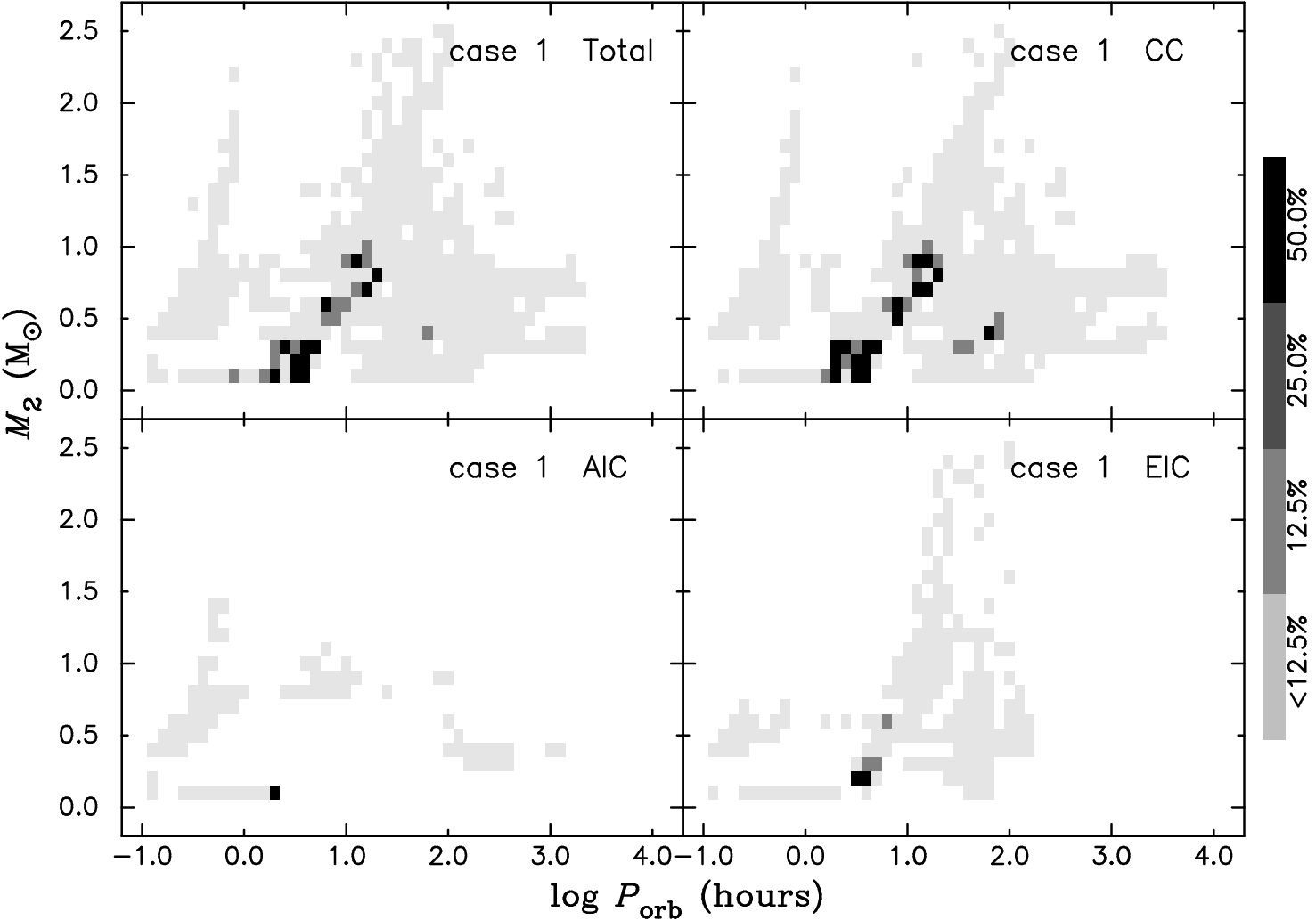}&
\includegraphics[totalheight=3.0in,width=2.5in,angle=-90]{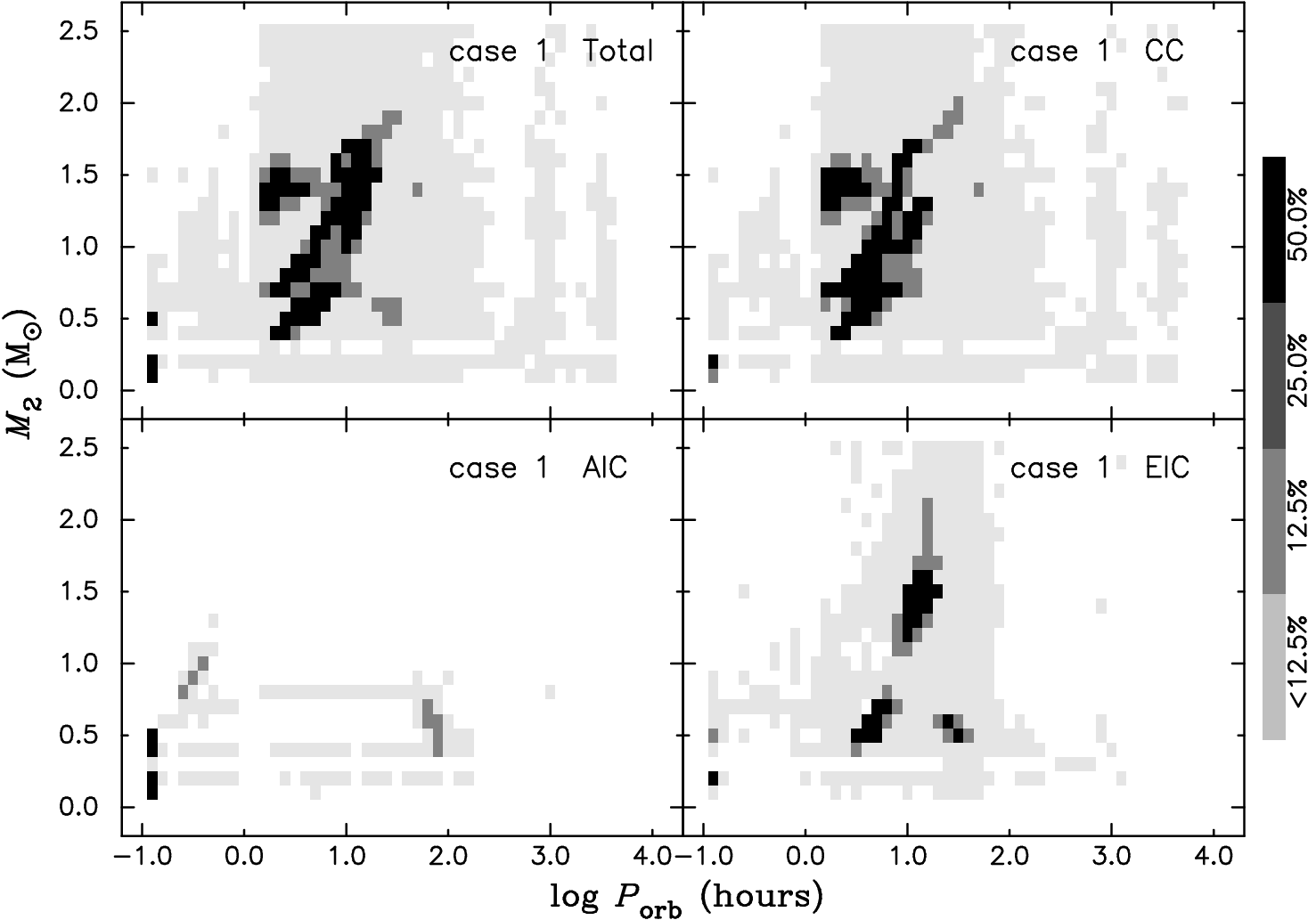}\\
\end{tabular}
\caption{Similar with Figure \ref{fig:im2po}, but for grey-scale maps of the distributions of the orbital periods $P_{\rm orb}$ vs.
the donor's masses in the MSXPs. The left is for the transient sources,
and the right is for the persistent sources. }
\label{fig:m2po}
\end{figure*}

Figure \ref{fig:m2po} gives the distributions of the donors' masses vs. orbital periods of the MSXPs.
Compared with Figure \ref{fig:im2po}, the progenitors of MSXPs have undergone violent mass transfer.
The distribution of the companion stars' masses for AIC
channel is narrower than those of for CC and EIC  channels. The interpretation for this distinction is that AIC binary
systems may have experienced significant accretion phases, and a significant amount of material has been
transferred to NSs, then the masses of companion stars are usually relatively small. In ultra-compact MSXPs,
the distribution of donors' masses are cut into two regions. In the upper region ($M_2> 0.4M_\odot$), the most of donors are naked helium stars
, while they are WDs in the down region.  For the MSXPs with orbital periods longer than 100 hours, donors mainly are red giant and
they have low masses ($\approx 0.5 M_\odot$) in order to avoid dynamical mass transfer.

\subsubsection{Spin Period: $P_{\rm s}$}
\begin{figure*}
\centering
\begin{tabular}{c@{\hspace{3pc}}c}
\includegraphics[totalheight=3.0in,width=2.5in,angle=-90]{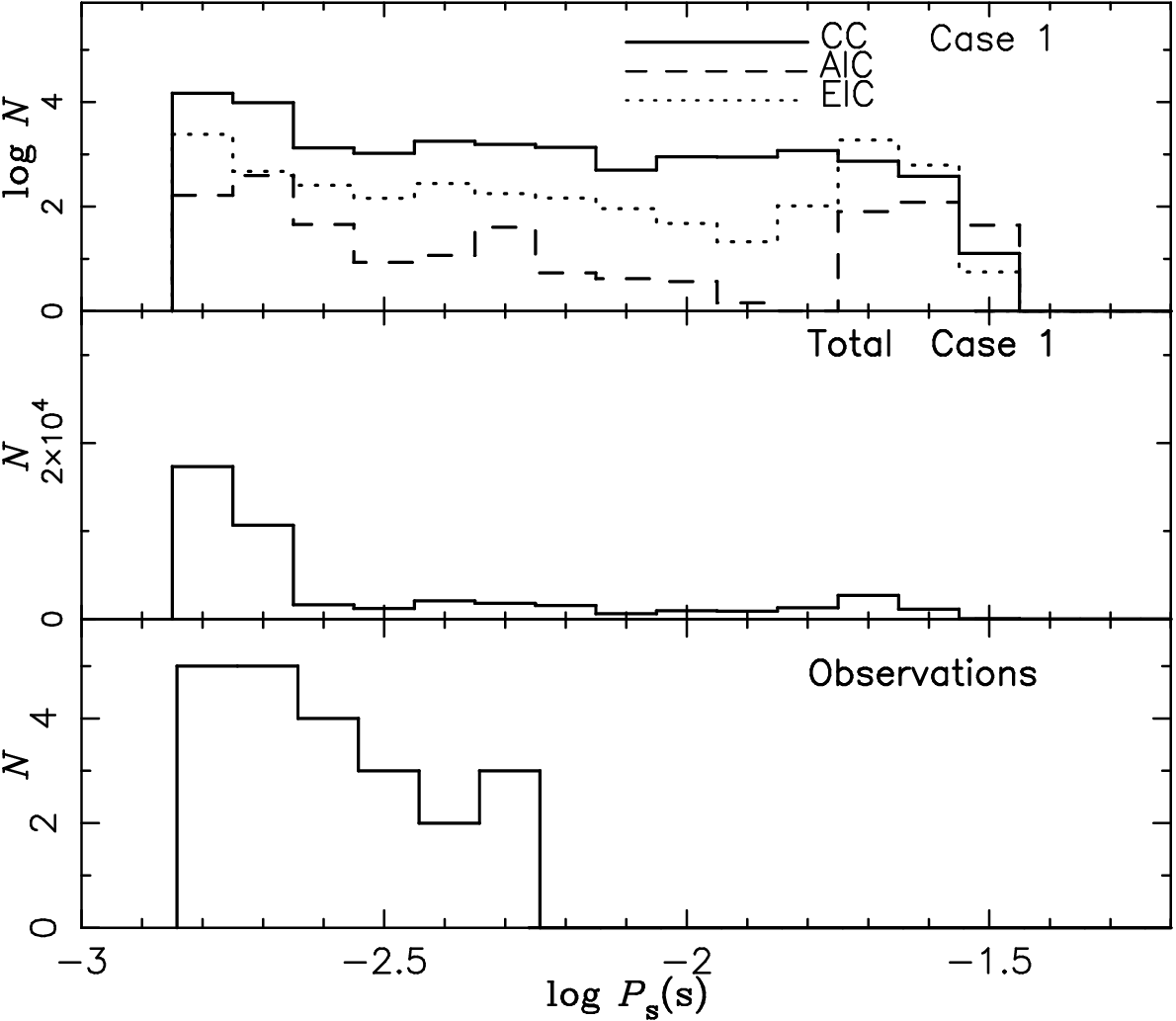}&
\includegraphics[totalheight=3.0in,width=2.5in,angle=-90]{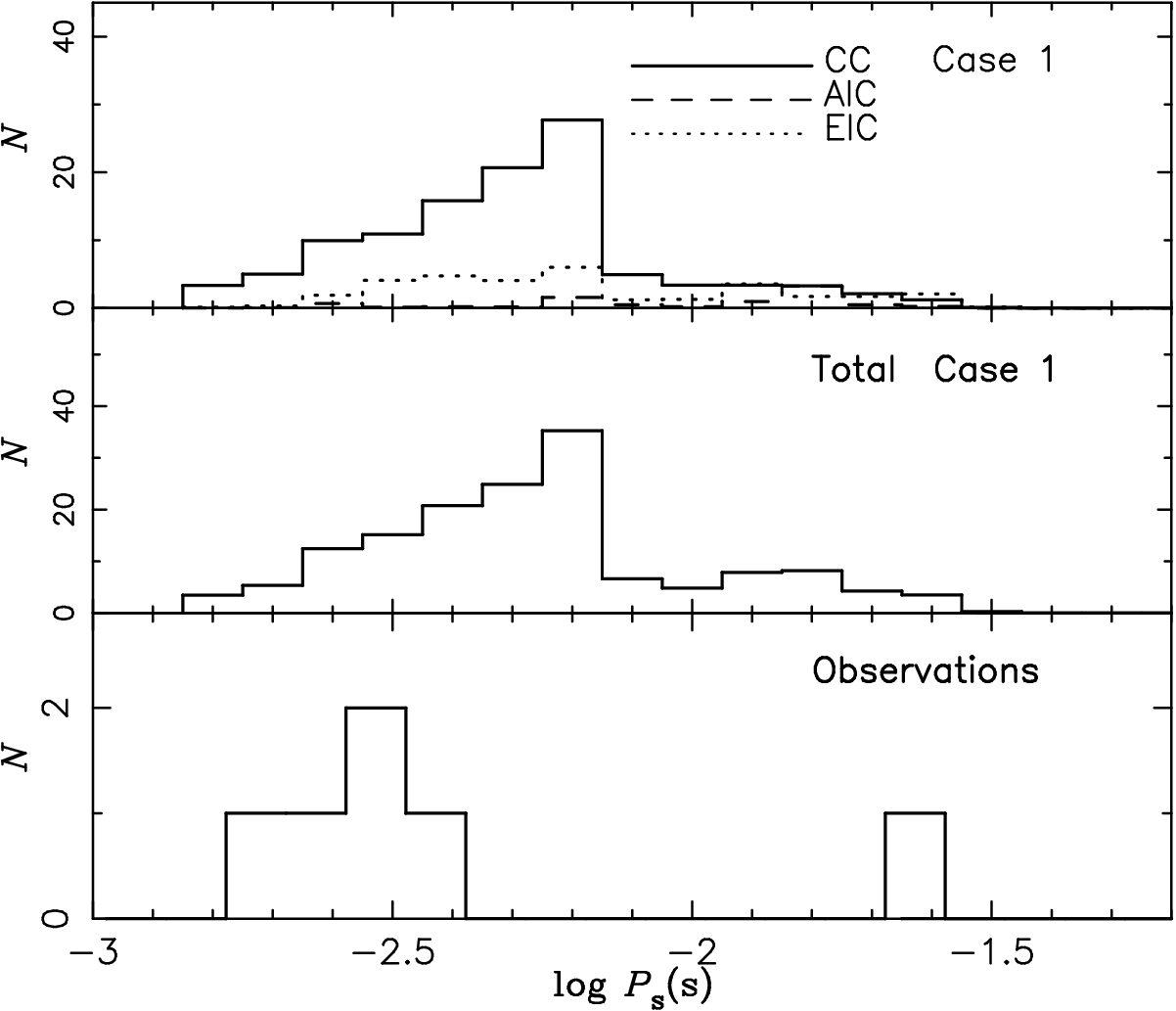}\\
\end{tabular}
\caption{The distributions of the spin periods of NSs in the MSXPs. The left is for the transient sources,
and the right is for the persistent sources. The data of observed MSXPs come from \citet{Watts2008, Patruno2012, Papitto2013}.}
\label{fig:spin}
\end{figure*}

The spin periods are one of the most important physical parameters for MSXPs.
Figure \ref{fig:spin} shows the distributions of the spin periods of NSs in MSXPs.
For the transient MSXPs (See the left graph in Figure \ref{fig:spin}), there
are two peaks. The left peak around about $ 1.6$ ms originates from the transient MSXPs with MS or red giant donors,
and the right peak around about 20 ms from  the transient MSXPs with WD donors. Usually, the mass transfer between two
degenerate binaries mainly is driven by gravitational radiation, and the mass transfer between
a degenerate star and a MS or red giant mainly is driven by stellar evolution \citep{Hurley2002}.
In our work, the time of the former occurring is much longer than that of the later occurring,
which means that the NSs in transient MSXPs with WD donors have lower magnetic fields because of the decay of
magnetic fields \citep{Lu2012}. A lower magnetic field results in a
shorter spin period.
For the distribution of spin periods of NSs in the persistent MSXPs (See the right graph in Figure \ref{fig:spin}),
there is a peak at $P_{\rm s}\approx 5.6$ ms. As Figure \ref{fig:spin} shows, the majority of NSs in the transient MSXPs
have spin periods between about 1.4 and 6.0 ms, while the spin periods of NSs in the persistent MSXPs mainly are
between about 6.0 and 20 ms. In our simulations, the transient MSXPs usually have shorter spin periods than those of
the persistent MSXPs. The main reason is that, compared with the NSs in the persistent MSXPs,
the NSs in the transient MSXPs have longer accretion histories which results
in lower magnetic fields.
As the bottom panels in Figure \ref{fig:spin} show, our results are in agreement with the observational samples.

\subsection{Corbet Diagram of MSXPs }
\begin{figure*}
\centering
\begin{tabular}{c@{\hspace{3pc}}c}
\includegraphics[totalheight=3.0in,width=4.5in,angle=-90]{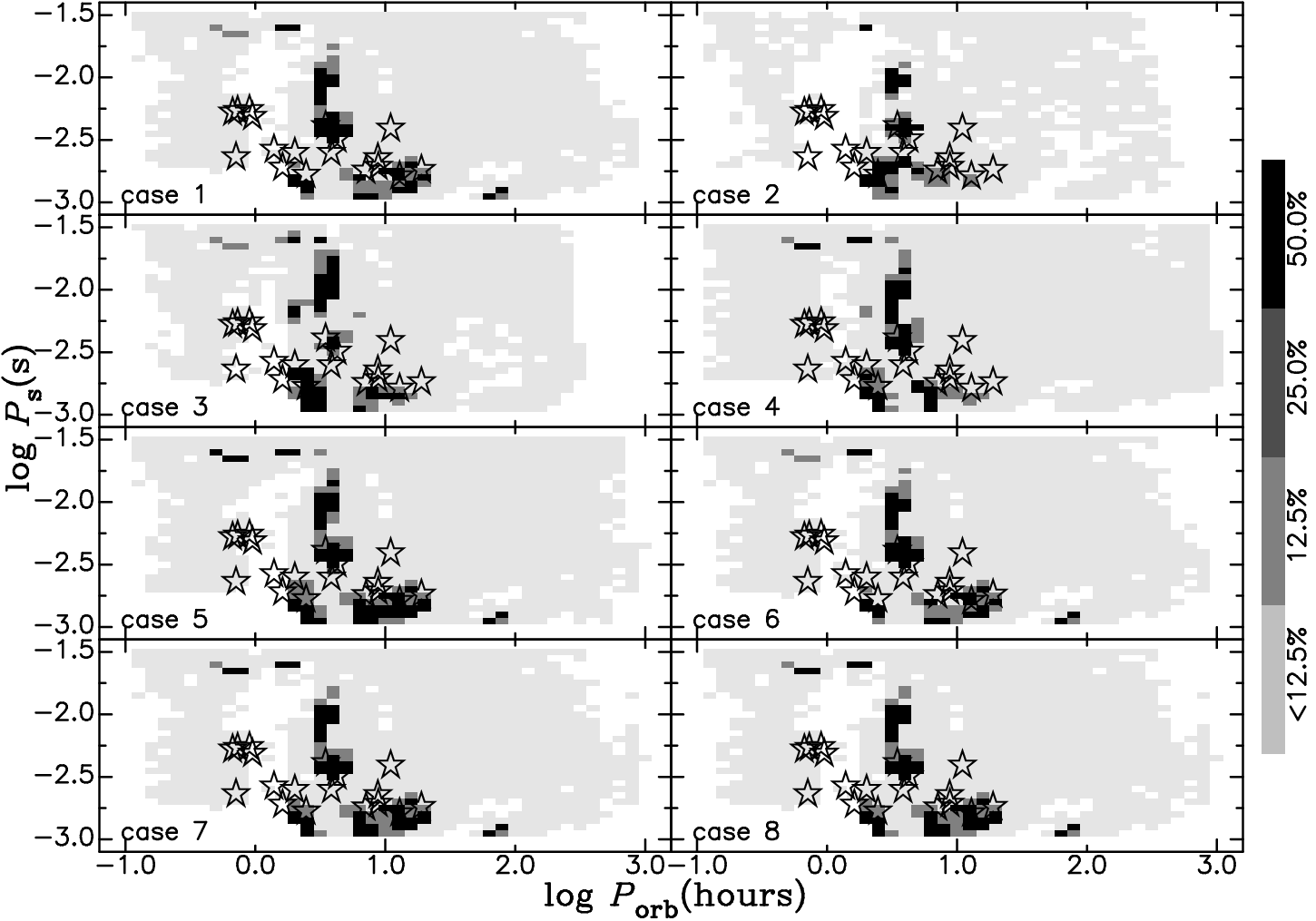}&
\includegraphics[totalheight=3.0in,width=4.5in,angle=-90]{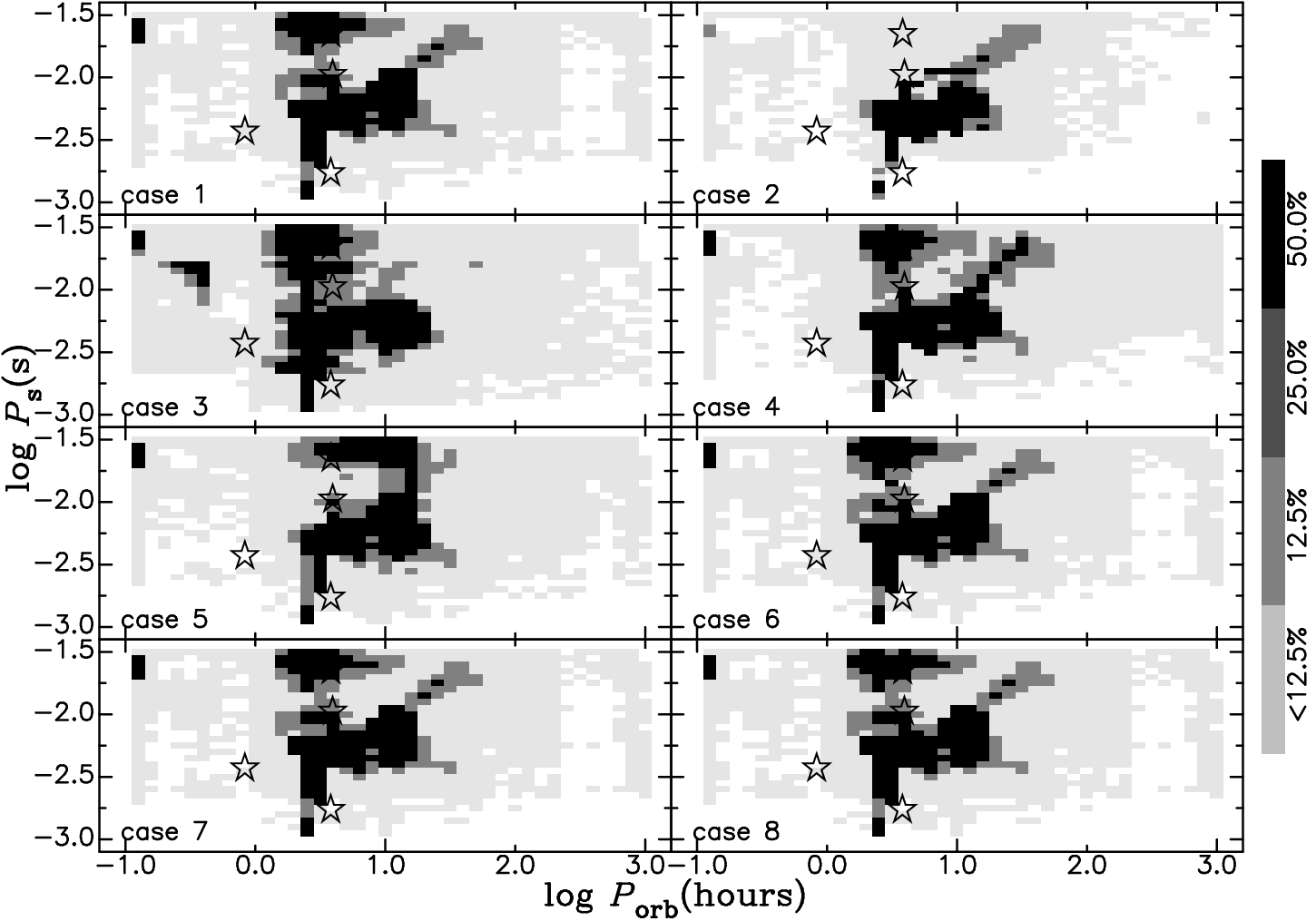}\\
\end{tabular}
\caption{Similar with Figure \ref{fig:im2po}, but for grey-scale maps of the distributions of the spin periods $P_{\rm s}$ vs.
the orbital periods $P_{\rm orb}$ for the MSXPs. The left is for the transient sources,
and the right is for the persistent sources. The data of observed MSXPs
\citep{Watts2008, Patruno2012, Papitto2013} are plotted by pentagram.}
\label{fig:pspo}
\end{figure*}

According to the distribution of spin periods $P_{\rm s}$ vs.
orbital periods $P_{\rm orb}$, \cite{Corbet1986} divided high-mass X-ray binaries into three classes.
Therefore, Corbet diagram is very important for us to know X-ray binaries.
In the known 31 MSXPs, there are 18 transient sources and 4 persistent sources
whose orbital periods and spin periods are measured.
They are plotted in Figure \ref{fig:pspo} in which the distributions of the spin periods $P_{\rm s}$ vs.
the orbital periods $P_{\rm orb}$ for the MSXPs simulated in our work are given in grey-scale.
As Figure \ref{fig:pspo} shows, whether observed or simulated MSXPs are cut into two zones: ultra-compact and
normal sources.
For the transient and persistent MSXPs,  the observed normal sources are covered by the simulated results in
all cases. We can not suggest which one is better. However, all simulations underestimate
the rotation speeds of the NSs in ultra-compact MSXPs. As the last section mentioned, in this work, the mass transfer between two
degenerate binaries mainly is driven by gravitational radiation. However, it is very difficult to
accurately simulate the mass transfer between two
degenerate binaries. Our models may underestimate the mass-transfer rate between WDs and NSs.
A high mass-transfer rate is favor for forming MSXPs. Therefore, underestimating the mass-transfer rate between WDs and NSs
is just the reason that we underestimate the ratio of the
ultra-compact MSXPs to total MSXPs.

\section{Conclusions}
MSXPs are thought to be the progenitors of radio MSPs. As the evolutionary link
between radio MSPs and NSs in LMXBs/IMXBs MSXPs have played an important role in
testing theories of pulsar formation and evolution.
In this paper, we have investigated in detail the MSXPs arising from three different evolutionary channels
(including the EIC  channel, which has almost been ignored in previous studies of radio MSPs),
using the population synthesis method to the binary evolutions.

Our simulations predict that the Galactic birthrates of the transient MSXPs and the persistent MSXPs are about 0.7---1.4
$\times 10^{-4} \rm yr^{-1}$. The observable number of the transient and
the persistent MSXPs in the Galaxy are about 4000---10000 and 300---500, respectively.
 We have discussed the properties
of the MSXPs, such as orbital period, spin period, masses of the companion stars, and mass-accretion
rates of the NSs. Our results about the distributions of spin periods and orbital periods for
MSXPs are similar to the observations.

Our population synthesis calculations have shown that about 53\%---86\% of the transient MSXPs have undergone CC channel,
about 14\%---42\% of them have undergone EIC  channel,
and the transient MSXPs via AIC channel are negligible. The proportions
of the persistent MSXPs via the above three channels are about 30\%---70\%, 10\%---40\% and 10\%---30\%, respectively.
According to the theories
of MSPs formation and evolution, the MSXPs will eventually evolve to the radio MSPs.
Our simulations show
that about 58\% of radio MSPs result from CC channel, 7\% of radio MSPs result from
AIC channel, and 35\% of radio MSPs result from EIC  channel. Therefore the EIC  channel is one
of the most important channels in the formation of MSXPs and radio MSPs, and cannot be
ignored in studies of  MSPs.

However, the investigation of MSXPs¡¯ population involves many uncertainties (See section 2). Although we try our best to choose commonly accepted paradigms, our results can not explain observations in many respects, such as:  an order of magnitude mismatch between observed and predicted theoretically numbers of MXBPs, and underestimating the spin periods of accreting NSs in ultra-compact MSXPs and their numbers. The possible reasons are as the follows: (i) erroneous treatment of binary evolution, including CE evolution, accreting NS evolution, and so on; (ii) incorrect lifetimes involving mass-transfer model and uncertain duty cycles of transient MSXPs; (iii)not considering the observational selection effects. In a short, there still is a long way to go for completely understanding MSXPs.

\section*{Acknowledgments}

This work was supported by
XinJiang Science Fund for Distinguished Young Scholars under Nos. 2014721015 and 2013721014, the National Natural Science Foundation
of China under Nos. 11473024, 11363005 and 11163005.

\bibliographystyle{mn2e}
\bibliography{lglmn,zch}
\label{lastpage}
\end{document}